\documentstyle[epsfig]{mn}

\def\figdir{.}  
\def\epsscale#1{\epsfxsize=#1\columnwidth}
\def\plotone#1{\par\centerline{\epsfbox{#1}}}

\def\eg{{e.g.\ }}
\def\etal{{ et al.~\/}}
\def\ie{{i.e.~\/}}
\def\mf{{Minkowski functionals}}

\title{Measuring Shapes of Galaxy Images I: Ellipticity 
and Orientation}
\author[Rahman \& Shandarin]
{Nurur Rahman and Sergei F.\ Shandarin\\
Department of Physics and Astronomy,
University of Kansas, Lawrence, KS 66045, USA;\\
nurur@kusmos.phsx.ukans.edu, sergei@ku.edu}
\date{}
\begin{document}
\def\dim#1{\mbox{\,#1}}
\maketitle

\begin{abstract}
We suggest a set of morphological measures that we believe can help
in quantifying the shapes of two-dimensional cosmological images such 
as galaxies, clusters, and superclusters of galaxies. The method employs 
non-parametric morphological descriptors known as the Minkowski 
functionals in combination with geometric moments widely used in the 
image analysis. For the purpose of visualization of the morphological
properties of image contour lines we introduce three auxiliary ellipses 
representing the vector and tensor \mf. We study the discreteness, 
seeing, and noise effects on elliptic contours as well as their 
morphological characteristics such as the ellipticity and orientation. 
In order to reduce the effect of noise we employ a technique of 
contour smoothing.
We test the method by studying simulated elliptic profiles of toy 
spheroidal galaxies ranging in ellipticity from E0 to E7. We then 
apply the method to real galaxies, including eight spheroidals, three 
disk spirals and one peculiar galaxy, as imaged in the near-infrared 
$K_s$-band (2.2 microns) with the Two Micron All Sky Survey (2MASS). 
The method is numerically very efficient and can be used in the study 
of hundreds of thousands images obtained in modern surveys. 
\end{abstract}
\section{Introduction}
The morphology of the objects such as galaxies, clusters, 
superclusters and voids of galaxies can provide important clues 
for understanding the past and present physical processes which 
play significant role in their  formation. Theoretical models of 
the structure formation in the universe are currently based on the 
hierarchical clustering scenario, the most essential component of 
which is the idea of merging. During the evolution of the universe,
small systems merge due to gravitational attraction resulting in
formation of  larger clumps. Merging of two or more galaxies is a
violent process that significantly disturbs the shape.
Hence the morphological study of galaxies and/or clusters of galaxies
at present and high redshifts may reveal important information about
the rate of merging at different redshifts  and thus put a stringent
constraints on the models of the structure formation.

The full morphological description of structures requires both 
topological and geometrical characteristics, and in general, is 
a formidable task. In practice, one would like
to have as much information as possible expressed in terms
of few meaningful and robust parameters as possible. This is, in 
principle, very difficult if not impossible to achieve.
We describe and test a new method designed to quantify the
shape of a two-dimensional image. We employ a set of morphological
measures known as Minkowski functionals (hereafter MFs). The
MFs (Minkowski 1903) have already been used for detection and
studies of possible non-Gaussianity in CMB maps
\cite{sch-gor98,novikov-etal99,hobson-etal99,novikov-etal00,shandarin02,shandarin-etal02}
and shapes of the images of simulated clusters of galaxies
\cite{beisbart00,beisbart-etal01a,beisbart-etal01b}.
A subset of MFs, known as the scalar MFs, has been used 
for studies of three-dimensional patterns in the large-scale 
distribution of galaxies
\cite{mecke-etal94,sch-buc97,kerscher-etal97,schmalzing-etal99,kerscher-etal01a,kerscher-etal01b,she-etal03}.

In the current study we develop a new set of quantities derived
from the two-dimensions: scalar, vector and several tensor MFs, and
apply them to simulated elliptic galaxies and to real galaxies as 
imaged in the 2MASS survey \cite{jarrett00,jarrett-etal00}.
At present we do not make any attempt to develop a new galaxy
classification. 
We wish to emphasize that the use of these shape descriptors could 
be viable along with the use of conventional structural parameters 
used in the galactic morphological analysis. In this paper, we 
concentrate on measuring the ellipticity and orientation of the 
images as a function of their sizes. 

It is worth noting that the MFs provide a non-parametric description 
of the images implying that no specific or prior assumptions are 
need to identify shapes. The technique is numerically very efficient 
and therefore is applicable to large data sets (\eg SDSS image 
catalogue).

The paper is organized as follows. In Sec. 2 we give a brief
introduction to the shape descriptors based on  MFs and 
geometrical moments. Sec. 3
describes the parameters derived from the MFs to quantify
shapes of simulated images. In Sec. 4 we study the effects
of discreteness, atmospheric seeing, and noise. In Sec. 5 
we describe a simple technique for contour smoothing
and show how it reduces the distortions caused by the noise
and thus improves the measurements of actual morphology.
Sec. 6 discusses a few examples of the images of real galaxies
from 2MASS survey. The conclusions of our investigation are
summarized in Sec. 7.

\section{Minkowski Functionals as shape descriptors \label{mfs}}

The MFs consist of a set of measures carrying both the geometric
(\eg areas, perimeters) and topological (the Euler characteristic)
information about the image. The MFs obey a set of covariance
properties such as motion invariance, additivity and continuity.
Here we use the scalar, vector and a selection of tensor MFs. The 
quantitative characteristics of images described below can be also 
characterized as geometric moments, \eg \cite{muk-ram98}. We use 
the so called silhouette and boundary moments up to second order. 

\subsection{Scalar MFs}
There are three scalar MFs derived from two dimensions: the area 
($A$), perimeter ($P$), and the Euler characteristic (EC, $\chi$)
of any region specified by a contour (\eg isophotal contour)
\begin{eqnarray}
A &=& \int\limits_{K} \ da, \label{A} \\
P &=& \oint \ dl,\\         \label{P}
\chi &=& \frac{1}{2 \pi} \ \oint \ \kappa \ dl,
\label{chi}
\end{eqnarray}
where $\kappa=1/R$ is the curvature of the contour, and $K$ is 
the region bounded by  a given contour. For all simply connected 
regions $\chi=1$ and therefore is not used in the paper.
Note also, that in cosmological studies of two 
dimensional fields the genus is often used instead of EC. The 
genus and EC are uniquely related and thus carry exactly the same 
information. The three scalar moments are translationally and 
rotationally invariant and thus do not carry any directional 
information. The scalar MFs can be considered as three geometric 
moments of zeroth order.
\subsection{Vector MFs}
Three vector MFs defining three centroids: the center of the 
area ($A_i$), the center of contour ($P_i$), and the center of 
curvature ($\chi _i$) are given as follows
\begin{eqnarray}
A_i &=& \frac{1}{A}\int\limits_{K} x_i \ da,  \label{vA}\\
P_i &=& \frac{1}{P}\oint \ x_i dl,  \label{vP} \\
\chi _i &=& \frac{1}{2 \pi} \ \oint \ x_i\kappa \ dl.
\label{vchi}
\end{eqnarray}

The vector MFs are in fact the center of mass ($A_i$) of the region 
within the contour assuming that the surface density is constant,
center of mass of the homogeneous contour ($P_i$), and center of mass
of the contour ($\chi _i$) having the linear density that equals the
curvature $\kappa$. Three centroids obviously coincide with each other
in the case of centrally
symmetric images but are generally different if the central symmetry
is broken. The centroids are the geometrical moments of the first order
carrying directional information.
\subsection{Tensor MFs}
Out of many tensor MFs \eg \cite{beisbart00} we use only
the following three central moments of the second order
\begin{eqnarray}
A_{ij} &=& \int\limits_{K} (x_i-A_i) (x_j-A_j) \ da, \label{tA}\\
P_{ij} &=& \oint \ (x_i-P_i) (x_j-P_j)dl, \label{tP} \\
\chi _{ij} &=& \frac{1}{2 \pi} \ \oint \ (x_i-\chi _i) (x_j-\chi _j)\kappa \ dl.
\label{tchi}
\end{eqnarray}

The tensors $A_{ij}, P_{ij}$ and $\chi_{ij}$ (also known as the 
curvature tensor) are closely related to the inertia tensors of a 
homogeneous region, a homogeneous contour, and a contour weighted 
by the curvature, $\kappa$, respectively. The three sets of moments 
provide the lowest order geometrical characteristics of the image 
including its orientation. 

\subsection{Ellipse}
Galaxy samples indicate environment dependent number density of 
galaxies of different morphologies: elliptic, spirals, and 
irregular/peculiar types. In morphological studies of galaxies, 
a conventional method is to approximate a galaxy shape by an 
ellipse \cite{car-met80}. Since the ellipse fitting technique is 
well known and widely used, to get a better feeling of the MFs we 
think it would be useful to begin with an analytical calculation 
of the MFs of an ellipse. Let us consider the equation of an 
ellipse centered at the origin 
\begin{equation}
\frac{x^2}{a^2} + \frac{y^2}{b^2} =1,
\end{equation}
where $a$ and $b$ are the semi-axes ($b\le a$).
Then, the scalar MFs are
\begin{equation}
A^{ell} = \pi a b,~~~  P^{ell} = 4 a E(e),~~~
\chi^{ell} = 1,  \label{EsMFs}
\end{equation}
where $e=\sqrt{1-b^2/a^2}$ is the eccentricity of the ellipse,
$E(e)=\int_0^{\pi/2}(1-e^2\sin^2{\psi})^{1/2}d\psi$ 
is the complete elliptic integral of the second kind.
Due to central symmetry of the ellipse its vector MFs coincide
with the center of symmetry of the ellipse: $A_i=0$, $P_i$=0, 
and $\chi_i=0$.

The eigen values of the tensor MFs are respectively
\begin{equation}
A^{ell}_{xx} = \frac{\pi}{4}ab^3, ~~~
A^{ell}_{yy} = \frac{\pi}{4}a^3b  \label{EaMFs}
\end{equation}

\begin{eqnarray}
P^{ell}_{xx} &=& \frac{4}{3}a^3\frac{(1-e^4)E(e)-(1-e^2)^2 K(e)}{e^2},
\nonumber \\
P^{ell}_{yy} &=& \frac{4}{3}a^3\frac{(1-e^2)K(e)-(1-2e^2) E(e)}{e^2},
\label{EpMFs}
\end{eqnarray}

\begin{equation}
\chi^{ell}_{xx} = \frac{b^3}{a+b}, ~~~
\chi^{ell}_{yy} = \frac{a^3}{a+b}.
\end{equation}
Here $K(e)=\int_0^{\pi/2}(1-e^2\sin^2{\psi})^{-1/2}d\psi$ 
is the complete elliptic integral of the first kind.

\section{Method}
We treat an image as a set of contour lines built from a pixelized
map. First, we construct a contour at every chosen level by using
the linear interpolation technique described in
\cite{shandarin-etal02}. The contour
is represented by an ordered set of points. In addition to the
coordinates of each contour point the angles between the adjacent
segments are also used as they carry information about the curvature
of the contour.

The measurements provide three scalar (eq. \ref{A} -  \ref{chi}),
three centroids (eq. \ref{vA} - \ref{vchi}) and a total nine 
components of the tensor MFs (eq. \ref{tA} -\ref{tchi}).
In order to make representation more homogeneous and intuitive we
transform the tensor MFs into pairs of parameters: one of which is
the area and the other is the perimeter of the ellipse having 
exactly the same tensor MFs.

\begin{figure*}
\epsscale{2.5}
\plotone{\figdir/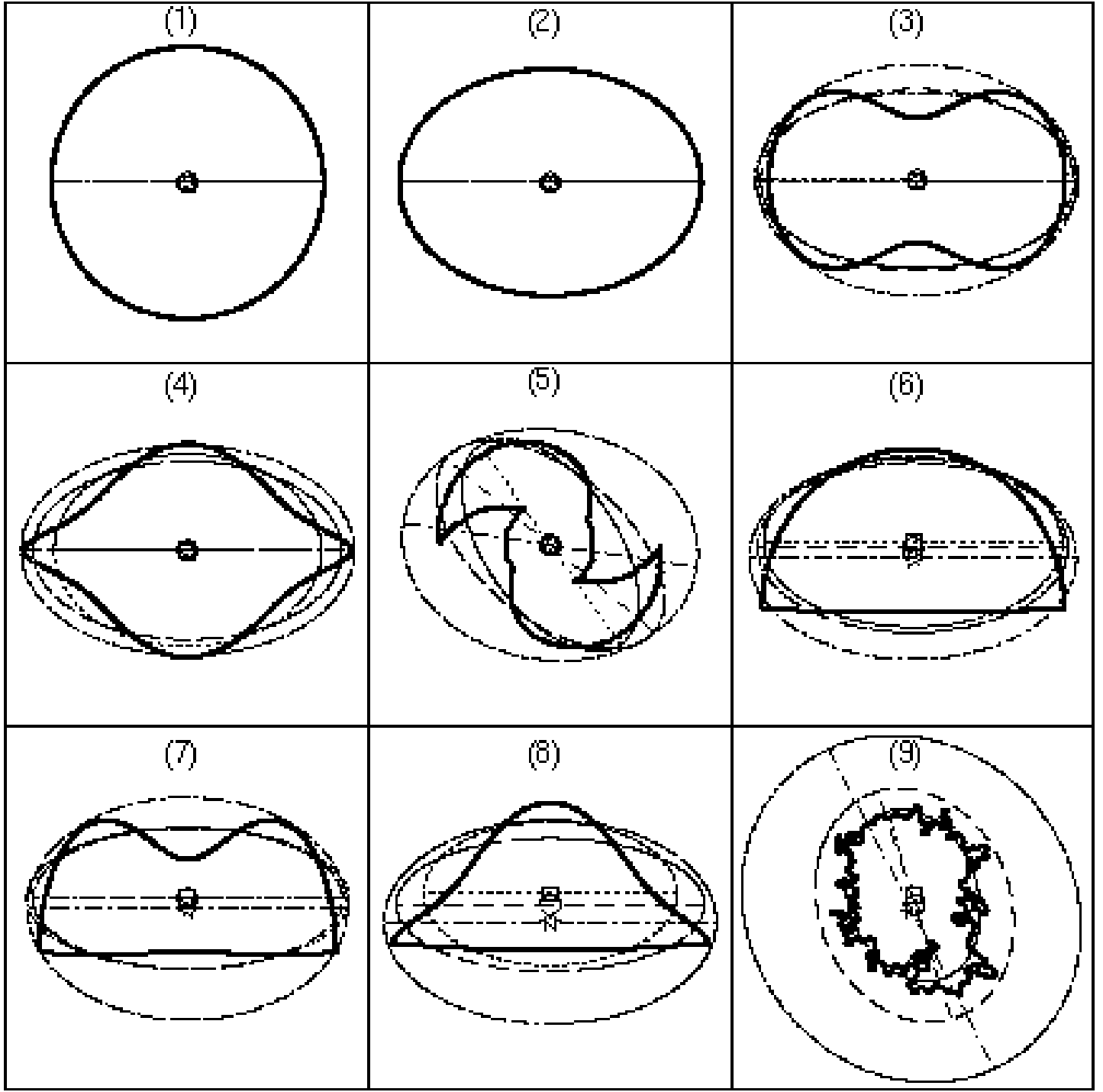}
\caption{A set of toy contours (solid lines)  with their vector
and tensor MFs. The vector functionals (centroids)
$A_i, P_i$, and $\chi_i$ are shown by the square, circle and 
star respectively. The tensor functionals are represented by the 
auxiliary ellipses with corresponding areas and perimeters: 
($A_A, P_A$ dotted), ($A_P, P_P$ dashed), and 
($A_{\chi}, P_{\chi}$ dashed-dotted). The straight lines passing 
through the centroids show the orientations of the corresponding 
auxiliary ellipses.
\label{toy_images}}
\end{figure*}

For instance, consider the tensor $P_{ij}$ (eq. \ref{tP}).
First, we find the eigen values of the tensor and insert them into
eqs. \ref{EpMFs}. Then, solving these equations for $a$ and $e$ we
find the parameters of an ellipse (\eg the semi-axes $a_P$, $b_P$)
having exactly same tensor $P_{ij}$ as obtained for
the contour in question. For brevity we shall call this
auxiliary ellipse the perimeter ellipse (compare to the image 
ellipses in \cite{muk-ram98}).
We characterize the perimeter ellipse by its area $A_P$ and
perimeter $P_P$ which are calculated by inserting $a_P$ and $b_P$
in eq. \ref{EsMFs}. Figure \ref{toy_images} illustrates this 
procedure: in every panel the solid line shows the contour while 
the dashed line shows the auxiliary perimeter ellipse having 
exactly the same $P_{ij}$ as the contour itself and the orientation 
of its largest axis is orthogonal to the orientation of the 
largest axis of the $P_{ij}$ tensor (eq. \ref{tP}).

Similarly the tensors $A_{ij}$ and $\chi_{ij}$
are represented by the area and curvature ellipses which are
characterized by the pairs  $A_A, P_A$ and
$A_{\chi}, P_{\chi}$ respectively and are shown in 
Fig. \ref{toy_images} by dotted and dashed-dotted lines.
The orientations of the auxiliary ellipses are marked by the
orientations of the corresponding largest axes in 
Fig. \ref{toy_images}.

Summarizing the previous description we present the list of the
parameters characterizing a contour
\begin{itemize}
\item three vectors $A_i, P_i$, and $\chi _i$ 
(eqs. \ref{vA} - \ref{vchi}) defining the centroids of the area, 
perimeter and curvature ellipses which coincide with the 
corresponding centroids of the region and contour themselves;
\item four areas:
the area within the contour itself $A_S$ (subscript $S$ indicates
that it is one of the scalar MFs) and
the areas of the auxiliary ellipses: $A_A$, $A_P$ and $A_{\chi}$;
\item four perimeters: the perimeter of the contour itself $P_S$ and
the perimeters of the auxiliary ellipses $P_A,P_P$ and $P_{\chi}$;
\item three angles showing the orientation of the three  auxiliary
ellipses.
\end{itemize}

In the ideal case of perfect measurements of a perfect ellipse all
the estimates would be the same:
the areas of all auxiliary ellipses are equal to the area of the
contour itself
($A_S=A_A=A_P=A_{\chi}=A_{cont}$) as well as  the perimeters
of the are equal to the auxiliary ellipses are equal to
the perimeter of the contour itself  ($P_S=P_A=P_P=P_{\chi}=P_{cont}$).
In addition, all three vectors
are the same ($A_i=P_i=\chi _i$) as well as the orientations
of all auxiliary ellipses. This is illustrated by panels 1 and 2 in
Fig. \ref{toy_images} where all three auxiliary ellipses are on top 
of each other and overlap with the contour itself.

However, if the contour is not elliptic, all the auxiliary ellipses
are in general different. The use of the auxiliary ellipses
effectively relates the contour to an ellipse: the similarity of three
auxiliary ellipses is a strong evidence that the shape of the contour
is elliptic. It is worth emphasizing that the plots of the auxiliary
ellipses are used only for illustrative purpose, the quantitative 
analysis of the images is based on the computation and comparison of 
their parameters. In our analysis we compute the parameters for a set 
of levels and plot them as a function of the area within the contour.

\subsection{Toy Examples}
In order to develop a better feeling how the parameters defined
in the previous section characterizing the shape of various 
contours we briefly discuss nine different contours shown in 
Fig. \ref{toy_images}.

In the first two panels all three auxiliary ellipses coincide with 
the contour itself indicating that the both contours (a circle in 
panel 1 and ellipse in panel 2) are true ellipses. In the seven 
remaining panels the auxiliary ellipses are all different. In most 
cases the area ellipse (dotted lines) is the smallest and the 
curvature ellipse (dashed-dotted lines) is the largest of the three.

The vector MFs mark the centers of the corresponding auxiliary 
ellipses and are shown by the open square, circle and star
corresponding to the area (eq. \ref{vA}), perimeter (eq. \ref{vP}) 
and EC (eq. \ref{vchi}) vectors MFs respectively.
If the contour has a center of symmetry then all three points
coincide with it (panels 1 through 5).
In the case of mirror symmetry of a contour they lie on
the axis of the symmetry (panels 6, 7 and 8). The mirror symmetry 
also results  in a similar orientation of all ellipses.

Finally, an asymmetric and irregular contour generally results in 
three auxiliary ellipses with different sizes and orientations.
This is illustrated in panel 9 which shows one real galaxy contour 
constructed form a 2MASS image of the peculiar galaxy NGC4485. 
Note that in addition to being classified as peculiar type, it is 
also classified as a barred, irregular, ``Magellanic'' type low 
surface brightness galaxy.

\section{Spurious Effects}
Here we show how discreteness, atmospheric seeing, and noise 
may affect the accuracy of the measurements of ellipticity and 
orientation of elliptic images.
\subsection{Discreteness}
\begin{figure*}
\epsscale{1.5}
\plotone{\figdir/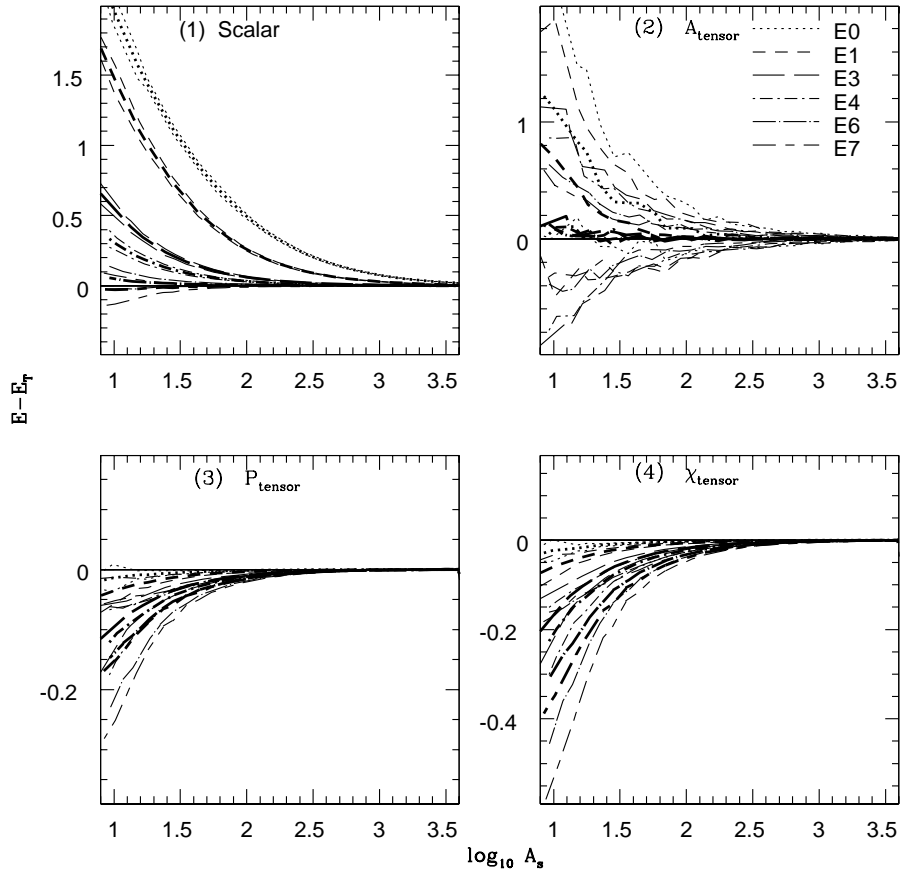}
\caption{The effect of discreteness on the ellipticity of 
elliptic contours. The horizontal axis is $\log_{10}A_S$ 
and the vertical axis is the errors in the measurements
of the ellipticity. Different lines mark ellipses with 
different ellipticities. Heavy lines correspond to the mean; 
thin lines show one sigma bands.
\label{discr}}
\end{figure*}

In order to assess the discreteness effect we generate one hundred
randomly oriented elliptic profiles (on a $128 \times 128$ mesh) 
with arbitrary position of the centers within a grid. We have 
compared the results of the measurements with analytic predictions 
in Fig. \ref{discr} where the error in estimated ellipticity is 
shown as a function of the area of the image. As one might expect 
the effect is stronger for smaller images and practically disappears 
for larger images. 
The four panels in this figure show the deviation of the measured 
ellipticity from the true one obtained from the scalar
($A_S, P_S$) and three tensor functionals. The ellipticity derived 
from the scalar functionals is systematically overestimated for all 
types of elliptic profiles (from E0 through E7) although the difference 
becomes smaller for elongated ellipses (E7). The estimate obtained from 
the area tensor ($A_{ij}$) has a relatively smaller systematic effect, 
however the dispersions remain quite large. 
The least distorted estimate of ellipticity comes from the perimeter 
($P_{ij}$) and the EC ($\chi_{ij}$) tensor. They both slightly 
underestimate the ellipticity. The elongated contours, e. g. E7, suffer 
more from discreteness than rounder one, e. g. E0, if derived from the 
perimeter and EC tensors while the opposite is true if derived from the 
scalar or area tensor. 
\subsection{Atmospheric Seeing and Discreteness}
\begin{figure*}
\epsscale{1.5}
\plotone{\figdir/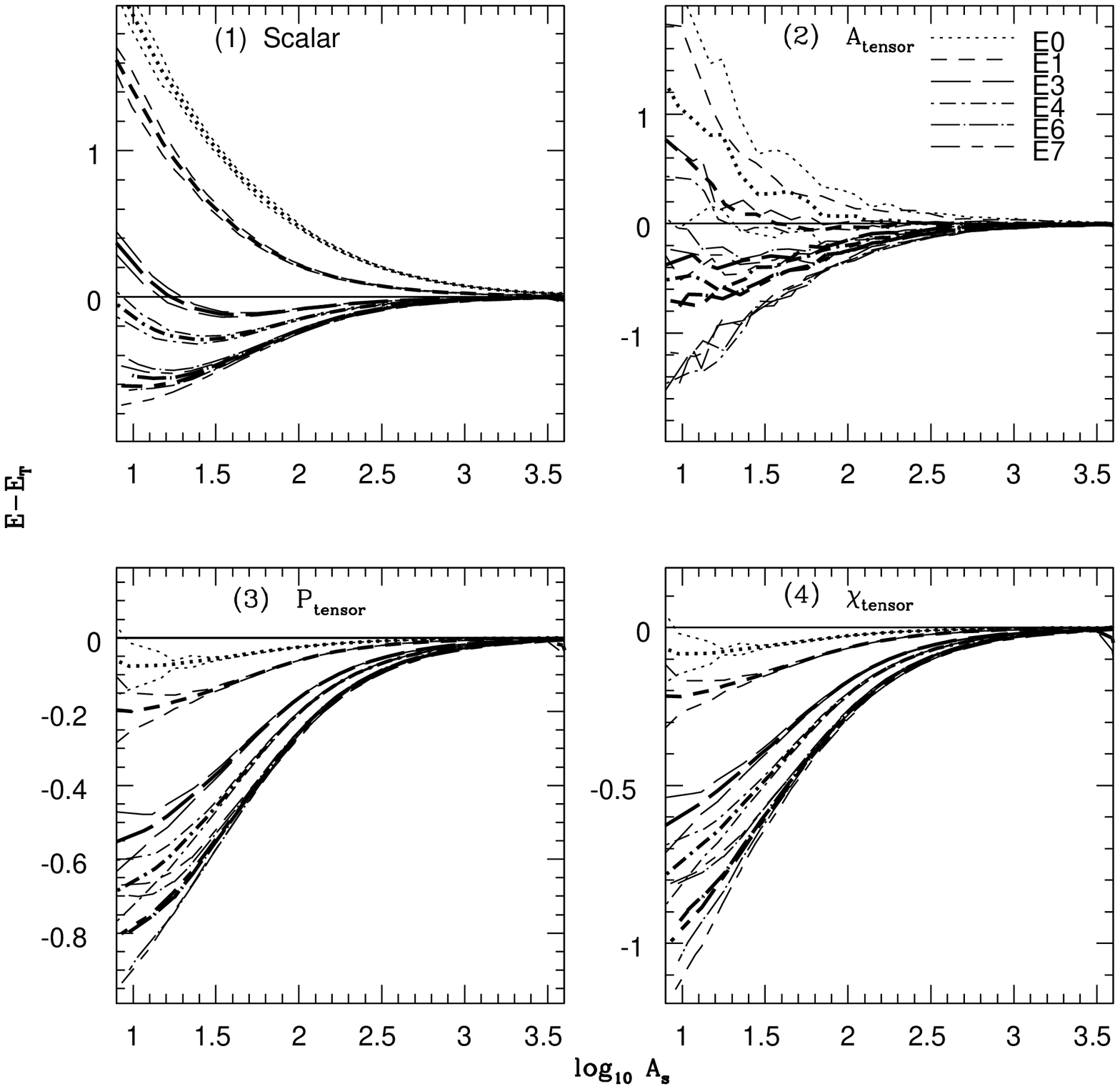}
\caption{The effect of atmospheric seeing and discreteness 
on the ellipticity  of elliptic contours. The profiles and 
notations are the same as in Fig. \ref{discr}. 
\label{seeing}}
\end{figure*}

An important effect that one must take into account while 
analyzing galaxy images is the atmospheric seeing. 
To quantify the effect of seeing on measured parameters 
we construct profiles as mentioned earlier and smooth them by  
2d Gaussian filter. The (real space) width of the filter is 
taken to be $\sim 1.5^{''}$, an approximate value of typical 
seeing of 2MASS observation although the actual PSF for the 
images ranges between $2.5^{''}$ and $3.^{''}$.
The profiles are analyzed as before and the results 
are shown in Fig. \ref{seeing}. Note that seeing is dominant 
around the central part of galaxy images which is also prone to 
the discreteness effect. Therefore interpreting the measurement 
in this region concentrating only on seeing will be erroneous 
and the understanding role of discreteness must also be needed. 
Figure \ref{seeing} shows their combined effect in ellipticity 
measurement of various simulated profiles.  

The seeing effect make the elliptical images appear rounder and 
thus reduces the ellipticity of the auxiliary ellipses for all 
profiles. It is interesting to note that the scalar estimator is 
slightly improved for rounder images (E0-E3) because the 
discreteness effects are partly compensated by the seeing. 
However, all other estimators are affected similarly to the 
discreteness effect and thus the ellipticity of the elliptic 
image becomes systematically smaller if estimated by the 
perimeter or the EC tensor. The area tensor still is the least 
affected by the seeing compared to the pure discreteness effect.
\subsection{Noise, Atmospheric Seeing, and Discreteness}
Noise is an additional source of distortion of the contours of any 
image. Originally smooth contours begin to wiggle due to noise 
affecting estimators derived from each functional. For instance, 
the perimeter and curvature systematically increase with the level 
of noise.
One may speculate that the parameters related to the curvature
(in our notations $A_{\chi}$, $P_{\chi}$) would be the most affected.
However, the EC itself (\ie $\chi$) does not change at all unless the
noise is so high that it breaks the continuity of the contour and 
split the initial region into two or connect it with another one.
The least affected MFs are related to the area and these are 
$A_S$, $A_A$, and $P_A$. 

\begin{figure*}
\epsscale{1.5}
\plotone{\figdir/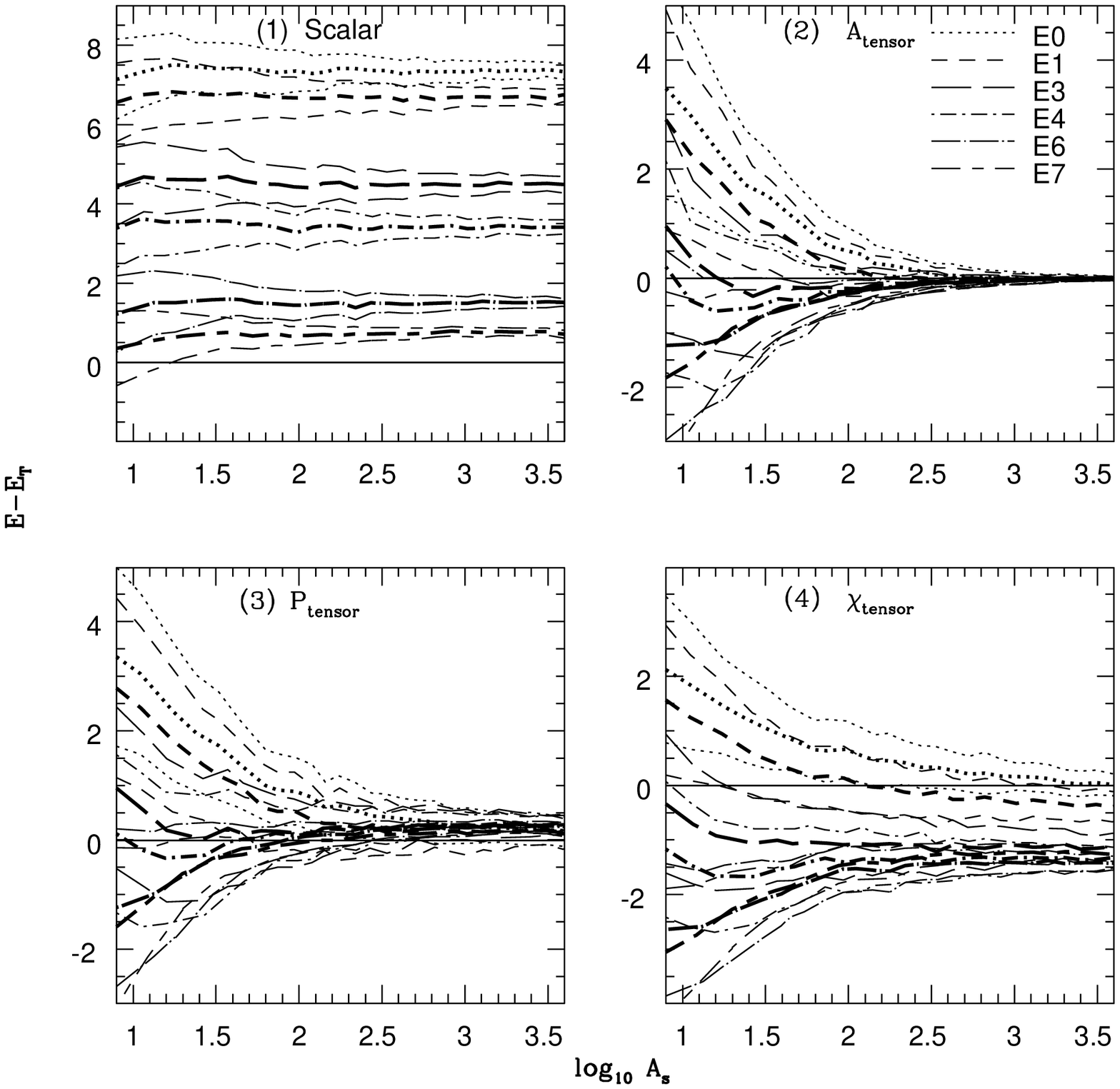}
\caption{The effect of noise, atmospheric seeing, and 
discreteness on the ellipticity of elliptic contours. The 
profiles and notations are the same as in Fig. \ref{discr}.
\label{noise}}
\end{figure*}

Figure \ref{noise} shows the combined effect of noise, seeing, and 
discreteness on the ellipticity of the image. The strongest 
systematic effect is on the scalar quantities (the top left panel).
The contour length becomes longer due to wiggling while 
the area is affected in less systematic way resulting a strong 
overestimate of the ellipticity, especially for rounder contours. 
Comparison of Fig. \ref{noise} to Fig. \ref{discr} and \ref{seeing} 
shows that the major effect is due to noise.

The effects of noise increase with decreasing signal to noise 
ratio S/N. A more detailed analysis shows that locally the
ratio of noise (the rms $\sigma_n$) to the gradient 
of the field ($\Delta f$) is the most important parameter 
controlling the distortions of the contour. For gradient we 
use $\Delta f =|\nabla f \cdot l_g|$ to be explicit in grid 
unit where  $l_g$ is the grid size. To visualize of what 
is stated above , consider an isophote corresponding to a given 
brightness level, $f_0$, passing through a particular grid point 
on a 2d mesh. As each grid point on the mesh has a particular 
brightness level, any other isophote belonging to a level higher 
or lower than $f_0$ will go through the inner or outer grids, 
respectively. In order for the isophote of $f_0$ to be shifted 
by the noise to a neighboring outward grid point, the level (or 
strength) of noise $\sigma_n$ must exceed the difference of 
unperturbed levels between the grid points. This shifting of 
the contour (i.e. distortion due to noise) solely depends on 
the gradient instead on the brightness level of the isophote. 

\begin{figure*}
\epsscale{1.1}
\plotone{\figdir/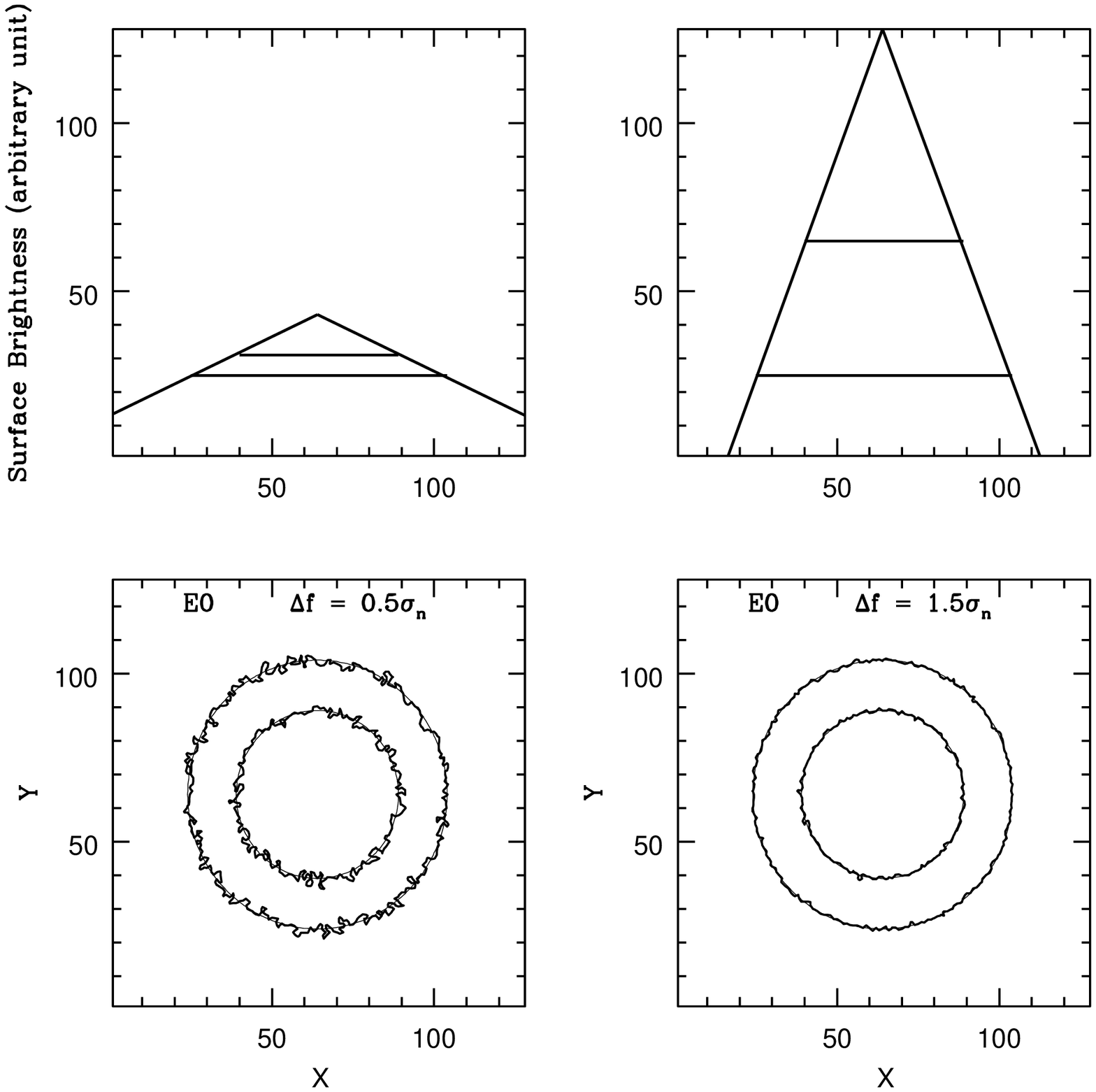}
\caption{A diagram showing the dependence between the surface 
brightness gradient and effect of noise. Top panels show 
un-perturbed 1d distribution curves for both low 
($\Delta f = 0.5\sigma_n$, top-left) and high surface brightness 
($\Delta f = 1.5\sigma_n$, top-right) profiles. 
The brightness curves are in arbitrary scale and the horizontal 
axis is in grid unit. The lower horizontal line show the same 
brightness level for both profiles  and the upper horizontal line 
is shown at different levels. These lines represent the diameters 
of the larger (80 grid unit) and smaller (50 grid unit) 
un-perturbed contours shown by thin solid in the bottom panels. 
The thick solid line is used to show noisy contours. For details 
see text. \label{cartoon}}
\end{figure*}

To elaborate more, we have included a diagram illustrating the 
dependence of gradient and image noise shown in Fig. \ref{cartoon}. 
Profiles with two different gradients ($\Delta f = 0.5\sigma_n$ 
and $\Delta f = 1.5\sigma_n$) are considered. The top panels 
show the un-perturbed 1d brightness distribution for lower 
(top-left) and higher (top-right) gradient profile. The brightness 
curves are shown in arbitrary unit and the horizontal axis is in 
grid unit. The longer horizontal line is at the same brightness 
level for both profiles where the shorter line is drawn at 
different levels. These lines represent the respective diameters 
(80 and 50, in grid unit) of the un-perturbed contours shown in 
the bottom panels. Each of the panels at the bottom contain both 
un-perturbed and noisy contours placed on top of each other. 
Thin and thick solid lines are used to show these respective 
contours. The outer contour is constructed by keeping the same 
size and brightness level for both profiles (lower horizontal 
line at the top panels). For the inner contour two different 
levels (the upper horizontal lines ) are chosen but keeping 
their sizes the same. 

With these choices made for contours, on one hand we see 
that the outer contour of the shallower profile is wiggled 
more and hence strongly distorted than that of the steeper 
one. The latter contour appear smooth and close in appearance 
to its original un-perturbed form. 
On the other hand we also notice from the appearance of the 
inner contour (for both profile) that the degree of distortion 
is not, at all, controlled by brightness level chosen. 
In spite of its higher level, the low gradient inner contour 
experiences similar distortion to the lower level outer 
contour. Contrary to this, the high gradient 
inner contour appear not only smooth but also similarly perturbed 
as its outer one. In brief, for low surface brightness profile 
the noise strength easily gains control over its steepness. Due 
to the intrinsic shape, the overall profile distribution experience 
strong influence of noise irrespective of the brightness level. 
For any given level, noise easily stretches the contour points 
along the x and y-direction on the 2d mesh and deviates the 
contours significantly from their original shapes. Comparatively 
speaking the shifts of the contour points on the mesh for a high 
brightness profile is less, resulting in smooth contours. 

It should be mentioned here that in the discussion above we 
considered an ideal situation assuming that the gradient 
remains constant all along the radial direction from center 
to the  edge of the profile. Since our goal is only to 
illustrate the significance of gradient-noise relationship, 
we believe this choice is reasonable.  
For real galaxy the steepness of brightness distribution 
varies radially from point to point, being steep at the 
central part and shallow outward. Due to this reason the 
real galaxy contours, in general, appear smoother in the 
center than at the edge.  
     
Any non-circular profiles have different gradients on a 
given contour at any given level and thus the amount of 
distortion varies along the contour. A natural assumption is 
that the average distortion of a contour of arbitrary shape 
is determined by the mean value of $\sigma_n/\Delta f$. 
For a linearized elliptic profile 
\begin{equation}
f_e(x,y)=F_e - g_e \sqrt{\frac{x^2}{a^2} +\frac{y^2}{b^2}}
\end{equation}
the mean inverse gradient can be evaluated analytically
\begin{equation}
\left \langle \frac{1}{|\nabla f_e|}\right \rangle 
= \frac{\pi b}{2 g_e E(e)}
\end{equation}
where $e$ is the eccentricity, $b$ is the semi-minor axis 
(in grid unit), $E(e)$ is the complete elliptic integral 
of the second kind, and $g_e$ is the gradient along the 
semi-major axis.

\begin{figure*}
\epsscale{1.1}
\plotone{\figdir/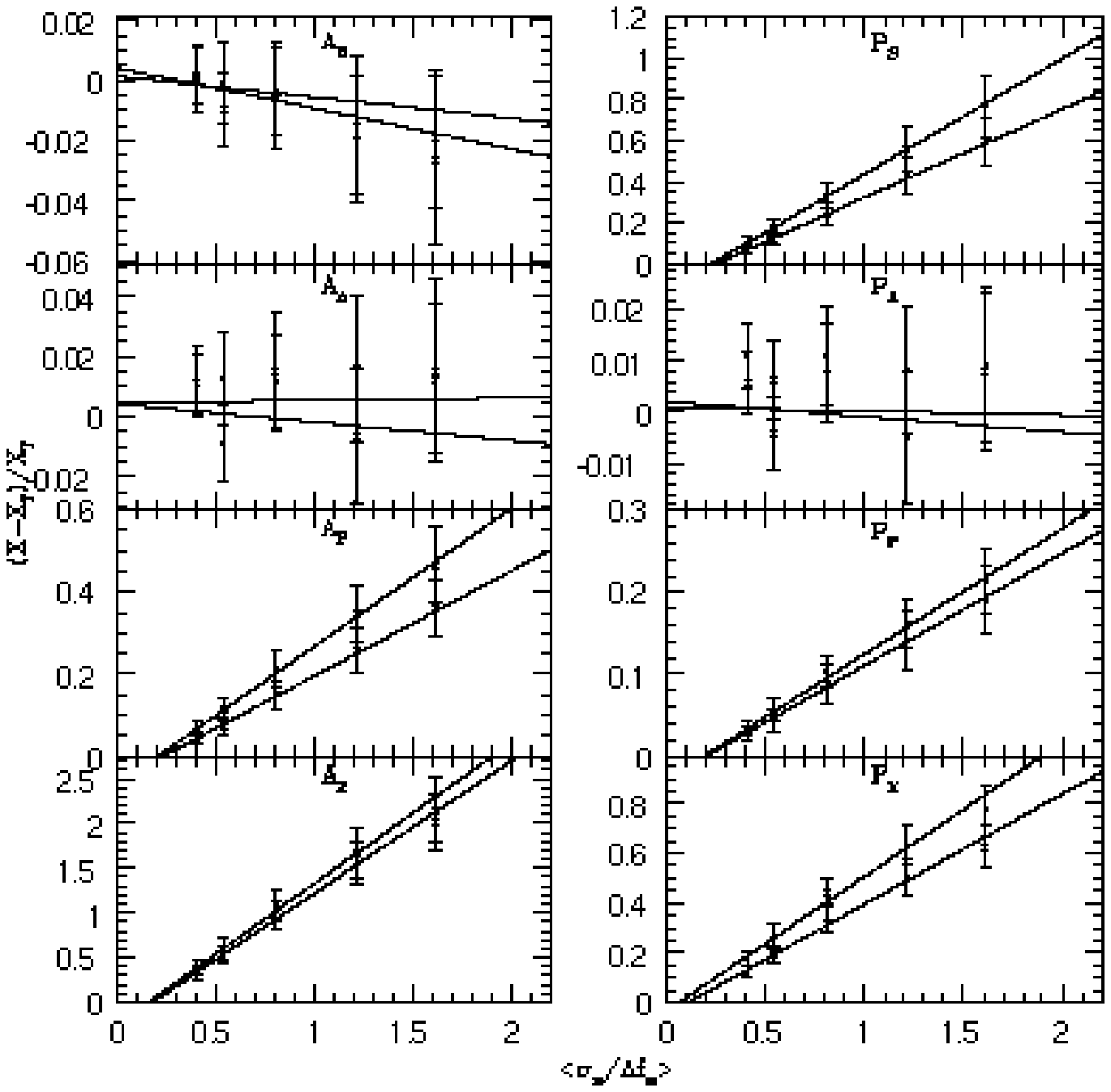}
\caption{The distortions of area and perimeter for E0 (thin l
ines) and E7 (thick lines) profiles are shown as a function of
$\left \langle \sigma _n / \Delta f_e \right \rangle$. The result 
shown here is obtained from a specific size contour of scalar 
area $A_S \approx 100$ although it is valid for any contour 
of arbitrary size. This choice for contour is made for ease of 
demonstration. \label{circ_ell_grad}}
\end{figure*}

A rounder and an elongated contour will be distorted 
similarly if they have the same mean inverse gradient. 
For an explicit demonstration and comparison of different 
profiles at approximately the same distortions, we have 
generated profiles with different ellipticities but similar 
slopes. Figure \ref{circ_ell_grad} illustrates the dependence 
of the distortions in areas (left panels) and  perimeters 
(right panels) for circular and elliptic profiles (E7) as a 
function of $ \left \langle \sigma_n / \Delta f_e\right \rangle$. 
For this illustration the images of a medium size 
$A_S \approx 100$ pixels have been chosen. The relative 
distortions ($(X-X_T)/X_T$ where $X$ is the measurement and 
$X_T$ is the theoretical, analytic in this case, value of 
the parameter in question) of the areas and perimeters are 
approximately linear with 
$\left \langle \sigma_n / \Delta f_e\right \rangle$ 
at least for $ \left \langle \sigma_n /\Delta f_e\right \rangle <2$
for both circular and elliptic profiles.
Moreover, the results for a circle and ellipse are approximately
similar, which confirms the assumption that the mean inverse 
gradient is one of the major factors affecting the amount of 
distortion. Note that only for theoretical model of elliptic 
profiles it is possible to express the gradient in closed 
analytic form and can be evaluated exactly. But in more complex 
cases of real galaxy profiles, analytic form is difficult 
to obtain and one must calculate the mean inverse gradient 
numerically.

\section{Contour Smoothing}
The effects of noise on image can be reduced in several possible 
ways. For example, the widely used method employ smoothing the 
image/map itself with some known filter function. In our 
analysis we chose a simple technique, known as the un-equally 
weighted moving average method, for contour smoothing.

The method is based on replacing the set of contour points by a 
new set each point of which is placed exactly in the middle of 
two adjacent points in the original set.
This set of new points make a new contour a little smoother than
the original one.  The procedure is applied
iteratively many times depending on the length of the contour
and the level of noise.
It is worth mentioning that finding the optimal number of smoothing
is not a trivial task.
The main problem is that without any prior knowledge of the true
shape one can in fact smooth a contour so much that
it will appear as circular and eventually with
further smoothing as a point since this is the ultimate situation.

One possible method of estimating the optimal amount of smoothing
could  be as follows.
One can make an image with known geometry (\eg elliptical peak or a 
real galaxy with high S/N ratio) and then
distort it by adding real noise taken from the nearby region of the
real image in question. Then applying the smoothing technique one 
can optimize its level on the reference image and use it for the 
study of the real image. However in this work we simply estimate 
the effect of Gaussian noise on elliptical profiles reserving a  
more elaborated technique for the future work.

The goal of smoothing is to restore the initial unperturbed 
contour as much as possible and measure its morphological 
parameters.
We simulate a hundred realizations of each profile positioning
the center randomly inside a pixel, rotating it by a random angle
and adding a realization of a Gaussian noise with a specified
level of $\sigma _n/\Delta f$. We measure the ellipticities for 
both distorted and smoothed contours.

\begin{figure*}
\epsscale{2.0}
\plotone{\figdir/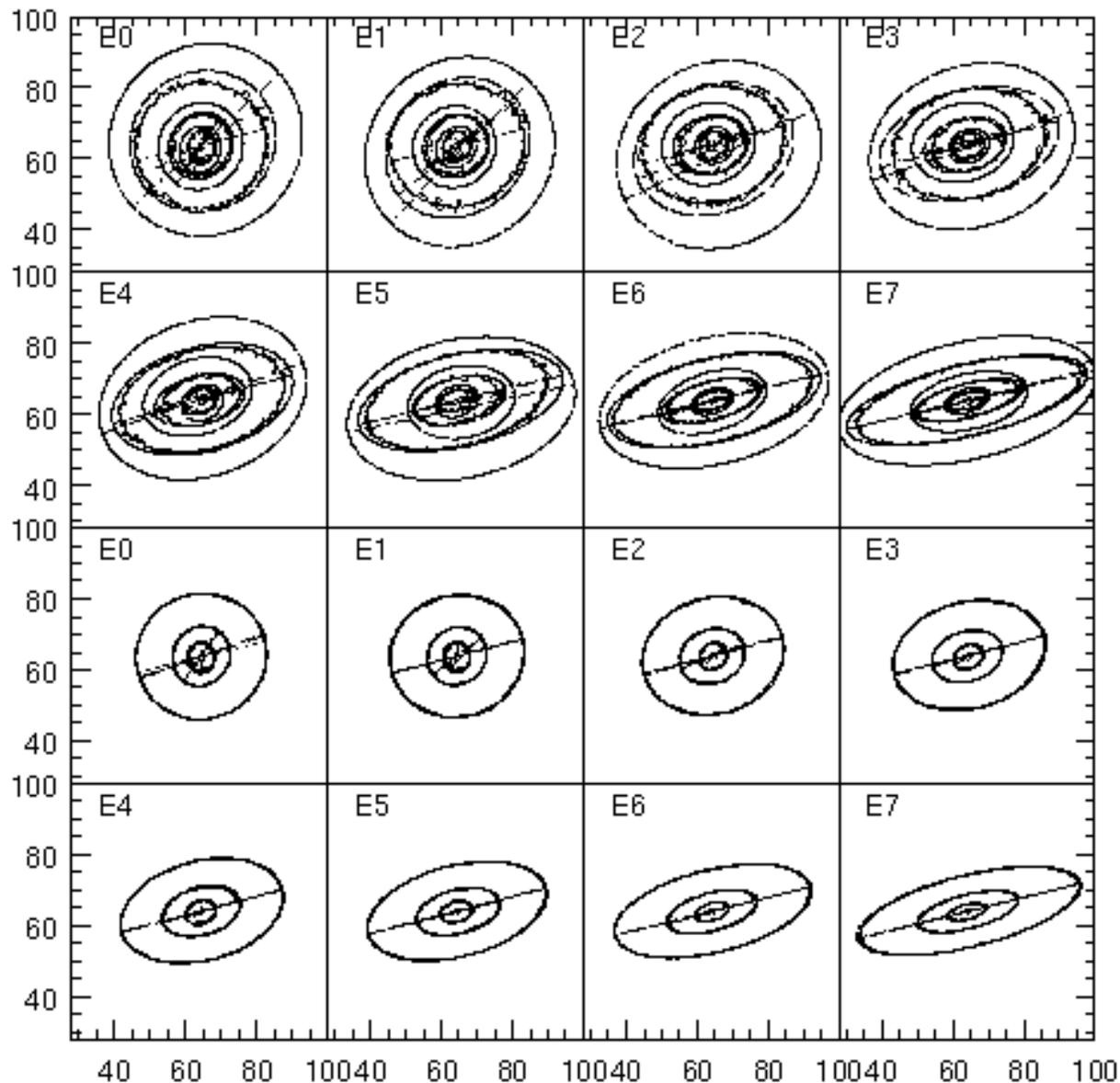}
\caption{Distortions of isophotal contours by seeing 
and Gaussian noise are shown for eight elliptical profiles.
The top two rows show three contours derived from the linear 
elliptical profiles with gradient $1.0 \sigma_n$ on the 
largest axis. Along with each contour three auxiliary 
ellipses are plotted (the line styles are similar as in Fig. 
\ref{toy_images}). Two bottom rows show the contours and the 
corresponding auxiliary ellipses after smoothing. The scalar 
areas of the contours are approximately $A_S \approx$ 50, 200, 
and 1000 in grid units.\label{8_ellipses}}
\end{figure*}

\begin{figure*}
\epsscale{1.5}
\plotone{\figdir/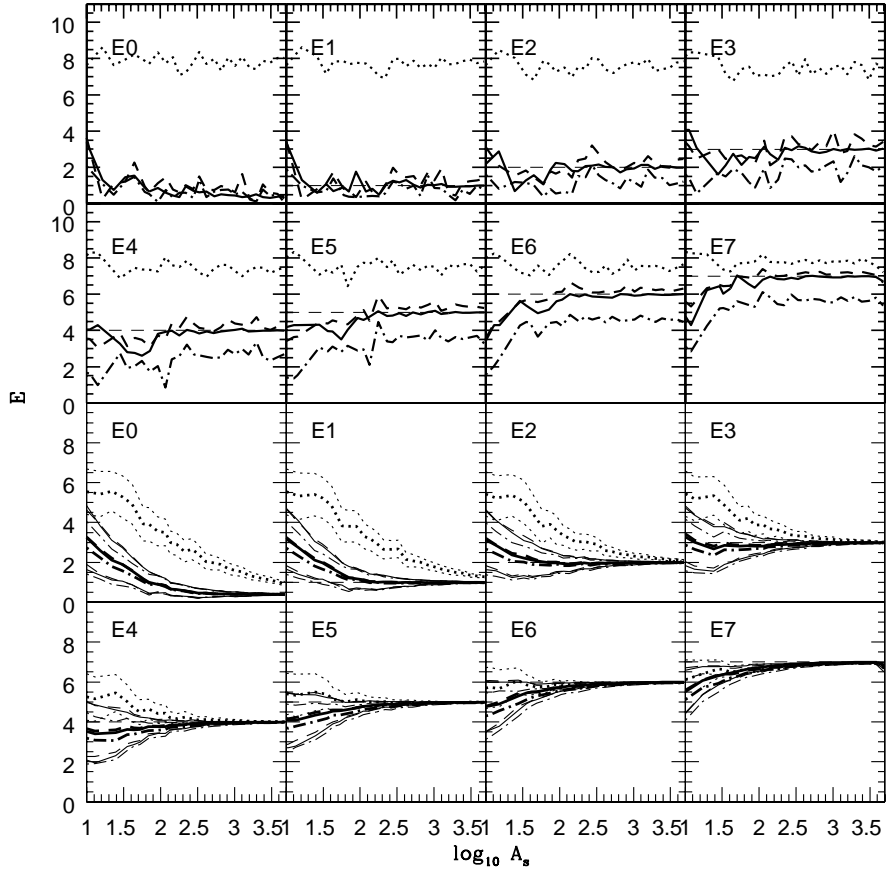}
\caption{The ellipticity parameter (E) for elliptical profiles 
shown in Fig.\ref{8_ellipses} before and after smoothing. The 
profile brightness distribution has gradient $1\sigma$ along 
the largest axis. The top two panels show the parameter measured 
from noisy profiles before smoothing. 
Result from only one realization is shown for demonstration 
purpose. The bottom two panels show mean and $1\sigma$ band for 
this parameters obtained from 100 realizations after smoothing.   
Horizontal dashed lines show the true ellipticity. The area, 
perimeter and curvature ellipses ($E_A$, $E_P$, and $E_{\chi}$) 
are shown by the solid, dashed, and dashed-dotted lines 
respectively. The dotted line shows the ellipticity of the 
scalar ellipse. Heavy lines correspond to the mean; thin lines 
show one sigma bands. 
\label{mfs_8_ellipses}}
\end{figure*}

\begin{figure*}
\epsscale{1.5}
\plotone{\figdir/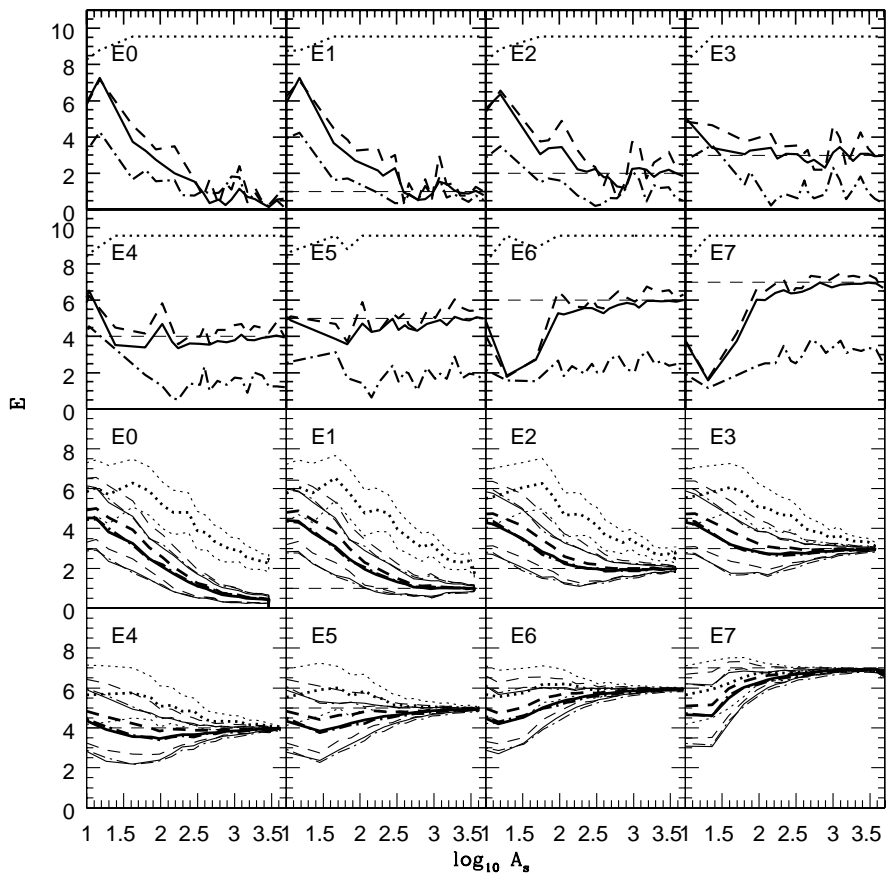}
\caption{Similar to Fig. \ref{mfs_8_ellipses} except for 
shallow brightness distribution. The profiles have gradient 
$0.2\sigma_n$ along the largest axis. This particular shallow 
distribution is chosen only to highlight the dependence of 
gradient and noise, and their roles on the measured parameter 
(E) as well as the significance of contour smoothing. Notice 
how distorted the profiles becomes owing to low brightness 
distribution (top two panels). Even though contour smoothing 
(in bottom two panels) substantially improve recovering shapes 
for larger elongated contours, for smaller and rounder contours 
(e.g. E0 - E2) it is harder to achieve. \label{mfs_8_ellipses2}}
\end{figure*}

The top two rows of panels of Fig. \ref{8_ellipses} show three 
selected contours for eight elliptical profiles (E0 - E7) 
distorted by Gaussian noise along with the auxiliary ellipses. 
The two bottom rows show the same contours after smoothing. 
In these rows the auxiliary ellipses practically coincide with 
the smoothed contours as expected. Figure \ref{mfs_8_ellipses} 
shows the results of measurement of ellipticities before (only 
for one realization) and after smoothing (from 100 realizations). 
The thin dashed lines show the true ellipticity of each profile. 
These figures show the 
fact that before smoothing different measures of ellipticity 
may give different results. The area tensor ($A_A, P_A$) gives 
the most accurate estimate of the ellipticity for all types of 
profiles. The estimate based on the perimeter tensor ($A_P, P_P$) 
closely follows it. The curvature tensor ($A_{\chi}, P_{\chi}$) 
is the least accurate estimator of ellipticity among all three 
tensorial measures; its accuracy decreases steadily with the 
increase of the ellipticity. The estimate based on the scalar 
\mf is strongly affected by the noise and is the least accurate 
for all but very elongated profiles (E6 - E7).  
Figure \ref{mfs_8_ellipses2} shows results for galaxy profiles 
with shallow brightness distributions.

\begin{figure*}
\epsscale{1.1}
\plotone{\figdir/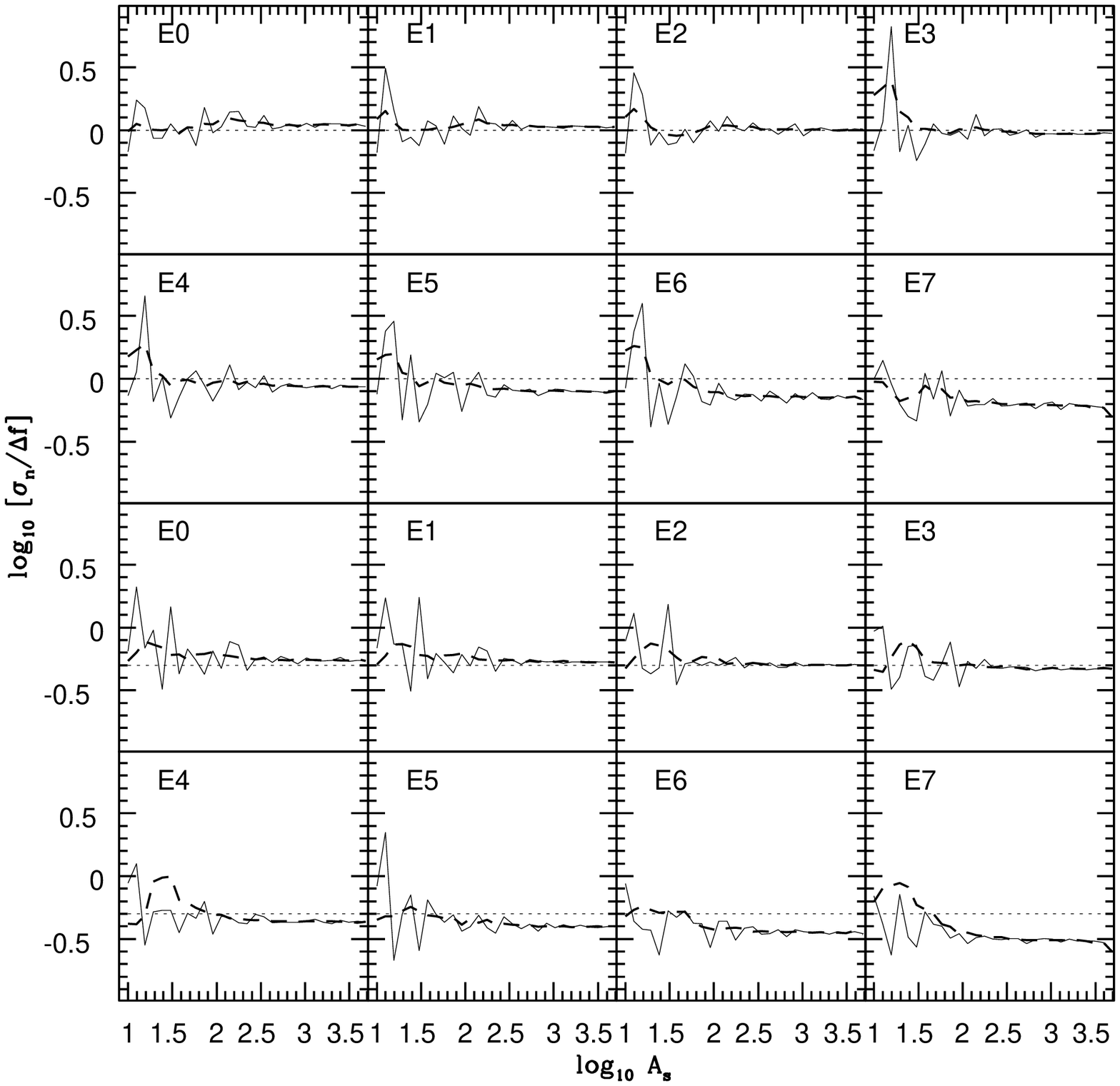}
\caption{The inverse gradient $\sigma_n/\Delta f$ as a 
function of log$_{10}A_S$ measured for eight elliptical profiles 
shown in Fig. \ref{8_ellipses}. The top two rows show the case
$\sigma _n /\Delta f =1$ and the bottom two rows correspond to
$\sigma _n /\Delta f  =0.5$. The dotted lines show the true value
of $\sigma _n / \Delta f$. \label{inv_g_ell}}
\end{figure*}

Contour smoothing results in a considerable improvement of the 
all estimates. As we mentioned before the number of smoothing 
steps is determined by the ratio $\sigma _n / \Delta f$ and 
thus we need to estimate it. We illustrate this using an example 
of circular contour. The gradient 
of a circular contour is given by $\Delta f/\Delta r$, where 
$\Delta f$ is the difference in brightness levels and 
$\Delta r$ is the difference in distances along the radial 
direction. One can rewrite the gradient as 
\begin{equation}
\frac{\Delta f}{\Delta r} = \frac{\Delta f}{\Delta A} \ 
\frac{\Delta A}{\Delta r}
\end{equation}
where $A$ is the area of the contour, $A = \pi r^2$ and $\Delta A$ 
is the difference in area of the corresponding to levels. With 
$\Delta A/\Delta r = 2\sqrt{\pi A}$, the equation for an estimate 
of inverse gradient boils down to the following expression
\begin{equation}
\tilde{G}  =  \frac{\tilde{G}_c}{\sqrt{A}} \ \frac{\Delta A}{\Delta f} 
\end{equation}
where \~{G} $= \Delta r/ \Delta f$, and 
\~{G}$_c$ $=1/(2\sqrt{\pi}) \approx 0.3$. 
Measuring \~{G} gives a reasonable empirical estimate 
of the inverse gradient $\sigma _n / \Delta f$. 
Figure \ref{inv_g_ell} shows the results of estimating
this quantity for two typical values of $\sigma _n /\Delta f$.
The dotted lines show the true value. The estimate of \~{G} 
for rounder contours (from E0 up to E3) overlaps nicely 
with the true value but becomes less accurate with 
increasing area as the contour becomes more elongated 
(from E4 and beyond). 
At larger area the $\Delta A$ is underestimated with the 
increasing flattening of the contour resulting a lower 
value ($\sim$ a factor of 2) of \~{G}. This corresponds 
to undersmoothing for larger flattend contours. Since the 
estimate of shape for these type of noisy contours show 
better result compared to rounder ones, small amount of 
smoothing is required for shape recovery and thus  
undersmoothing is compensated.  
The measurement of \~{G} can be improved if it is 
calculated for an elliptic contour instead of a circle. 
We reserve this issue to implement for our future work. 
The measured \~{G} oscillates at smaller area due to 
discreteness effect that eventually affects the final contour 
smoothing. This spurious effect is reduced by smoothing 
\~{G} by equally-weighted five points interpolation method. 
The number of smoothing steps ($N_s$) is finally achieved by 
multiplying the smoothed inverse gradient with the contour points 
($N_p$) at each level, i. e., $N_s = $\~{G} $N_p$. The $N_s$ acts 
as the upper limit of the iterative process (an un-equally weighted 
moving average method) through which contours get smoothed.
Note that the number of steps used to smooth galaxy 
contours increase from inner smaller regions towards the outer 
bigger regions. This is well justified by the fact that in the 
inner region a galaxy has lower inverse gradient (steeper slope) 
then the outer region (shallower slope). Contours are distorted 
heavily in the outer region by the noise, traverse longer paths 
consisting of large number of points and hence invoke higher 
number of smoothing steps. 

All estimates of ellipticity based on tensor functionals converge 
to one another after smoothing. For sufficiently large images 
($\log_{10}A_S > 1.8,~ A_S > 60~$ pixel) they also converge to the 
true values marked by the dashed lines in Figs. \ref{mfs_8_ellipses} 
and \ref{mfs_8_ellipses2}, respectively . 
The estimate of ellipticity based 
on the scalar functional ($A_S, P_S$) converge to the true value 
only for sufficiently elongated ellipses (E $>$ 2.5); it is worth 
noting that the larger the image the faster the convergence. The 
ellipticity of elongated contours (E $\ge$ 4) is estimated  more 
accurately by the scalar functional then by the tensors. Note that 
an accurate measurement of the ellipticity of circular contours is 
the most difficult to achieve.

\section{Examples of Galaxy Images}
In order to illustrate the technique we have taken several 
Near-Infra Red (NIR) images from 2MASS catalogue. These images 
are $101^{''} \times 101^{''}$ in size where $1^{''}/$pixel. The 
seeing FWHM values for 2MASS are typically in between $2.5^{''}$ 
and $3.0^{''}$. For details of 2MASS observation and data 
reduction readers are referred to \cite{jarrett-etal00,jarrett-etal03}.

\subsection{Elliptic Galaxies}
\begin{figure*}
\epsscale{2.0}
\plotone{\figdir/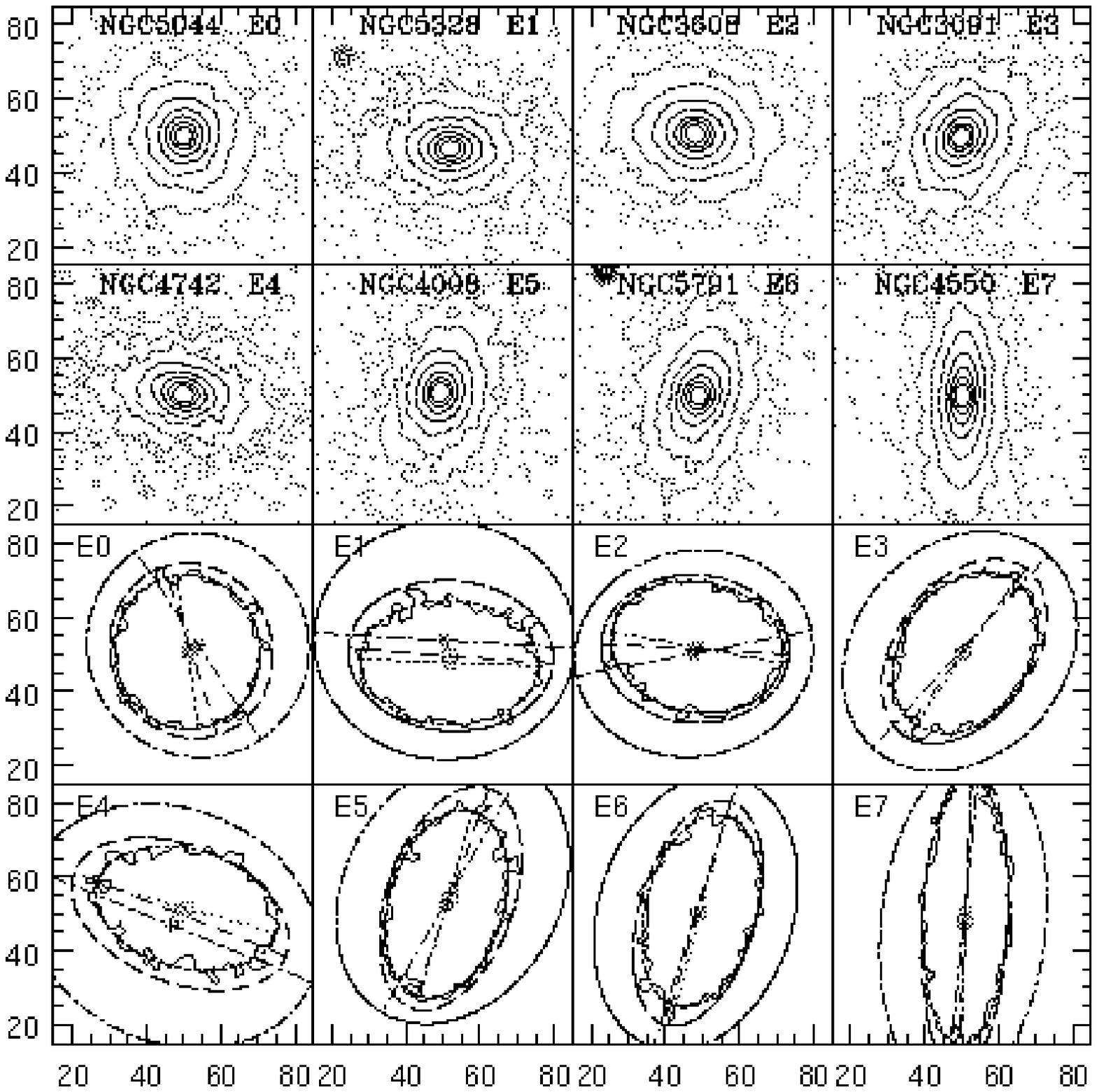}
\caption{Eight $K_s$ band galaxy images from 2MASS catalogue
(optically classified as ellipticals) are shown in  two top 
rows. Nine contours correspond to the areas $20\times 2^{n-1}$
in grid units ($n=1,...,9$). Two bottom rows show the contours 
corresponding to the area $A_S \approx 1300$ along with the 
centroids and auxiliary ellipses marked similarly as in 
Fig. \ref{toy_images}.
\label{8gal_im}}
\end{figure*}

In this present work we only consider the Ks-band (2.2 micron) 
images since they are less affected by the deleterious effects of 
atmospheric variability (the so-called "airglow" emission) that 
plague ground-based NIR observations."
We choose galaxies of different Hubble types such as NGC5044 (E0),
NGC5328 (E1), NGC3608 (E2), NGC3091 (E3), NGC4742 (E4), NGC4008
(E5), NGC5791 (E6), and NGC4550 (E7). 
Note that the classifications for the galaxies come from 
optical studies (primarily the RC3, de Vaucouleurs et al. 1991) 
due to the relatively insensitive separation of Hubble-types 
at near-infrared wavelengths (Jarrett 2000; Jarrett et al.2003).
The contour of these galaxies having equal area 
($A_S \approx 1300$) are shown at the bottom panels in 
Fig. \ref{8gal_im} along with the vector MFs and three 
auxiliary (tensor) ellipses.

\begin{figure*}
\epsscale{1.1}
\plotone{\figdir/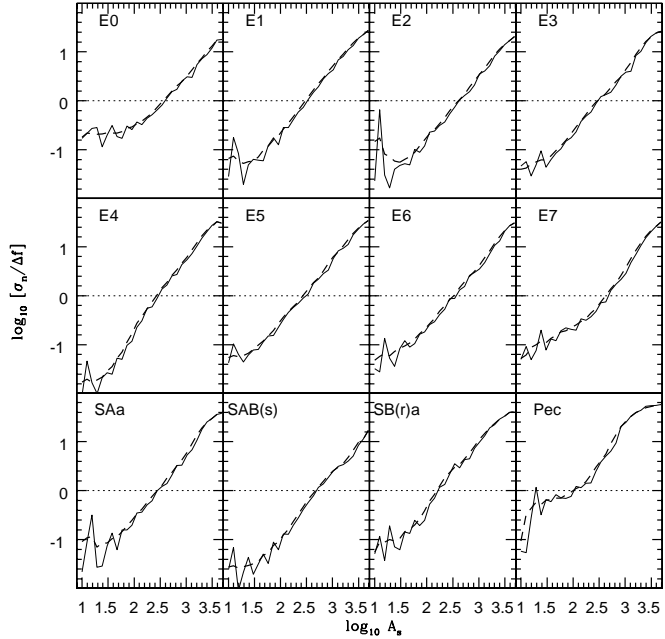}
\caption{The inverse gradient $\sigma_n/\Delta f$ as a function 
of log$_{10}A_S$ measured for eight galaxies from 2MASS catalogue 
(Fig. \ref{8gal_im}). The solid and dashed lines correspond to 
the raw and smoothed images respectively. The dotted line marks 
$\sigma_n/\Delta f=1$. \label{inv_g_gal}}
\end{figure*}

\begin{figure*}
\epsscale{2.0}
\plotone{\figdir/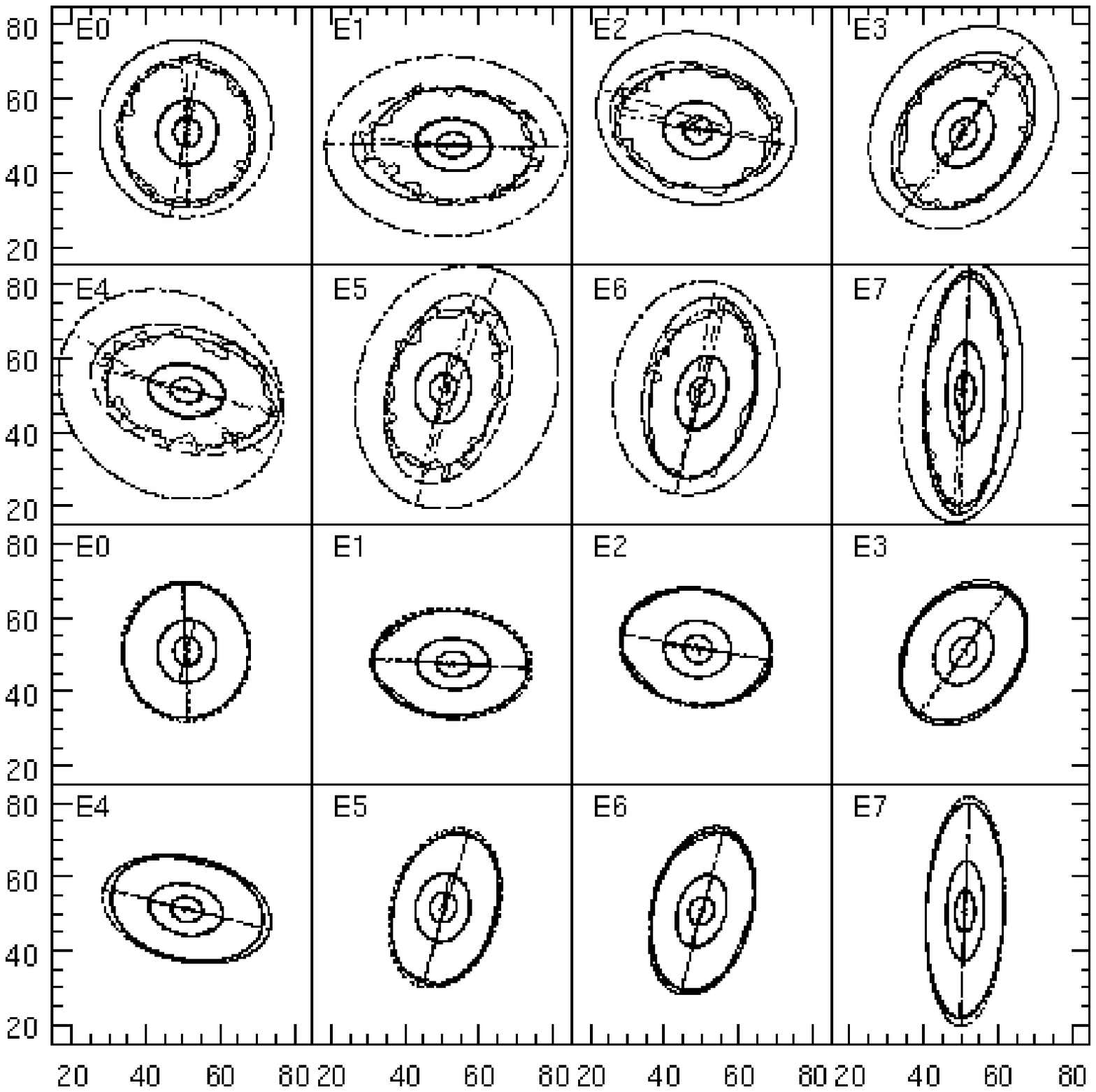}
\caption{The contours of eight galaxy images from 2MASS catalogue 
shown in Fig. \ref{8gal_im}. Three contours corresponding 
to the areas $A_S \approx $50, 200, and 1000 in grid units. 
Two bottom rows show the same contours after smoothing. 
The lines and legends are similar as in Fig. \ref{toy_images}.
\label{8gal_im_sm}}
\end{figure*}

To analyze the galaxies we first measure the ratio of 
$\sigma _n /\Delta f$ that will allow us to adjust the number 
of smoothing. Figure \ref{inv_g_gal} shows the results of the 
measurements: solid lines show the estimates before smoothing 
and the dashed lines after smoothing.
Figure \ref{8gal_im_sm} show three contours for each galaxy
before (top two rows) and after smoothing (bottom two rows).
The auxiliary ellipses for each contour are shown as well.

\begin{figure*}
\epsscale{1.1}
\plotone{\figdir/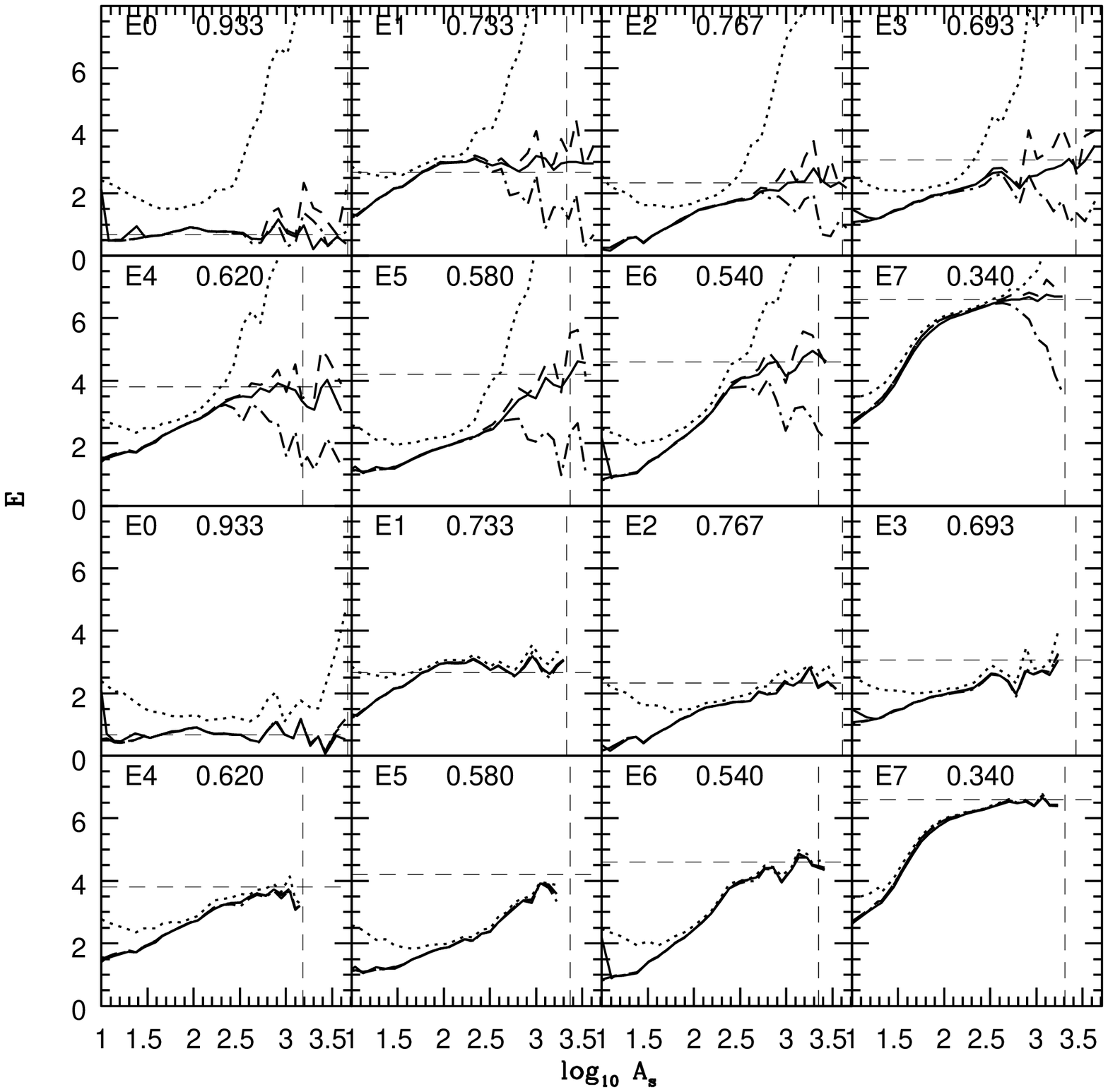}
\caption{The ellipticity of eight galaxies shown in 
Fig. \ref{8gal_im} as a function of log$_{10}A_S$. The 
ellipticities of the area, perimeter, and curvature ellipses 
are shown along with the ellipticity of the scalar ellipse. The 
line styles are similar as in  Fig. \ref{mfs_8_ellipses}. 
The raw images have been used in the top two rows. The bottom two 
rows show the ellipticities of smoothed contours. The horizontal 
dashed lines mark the $3\sigma_n$ isophote ellipticities of these 
galaxies given by the 2MASS catalogue. The vertical dashed lines 
show the scalar areas in our measurement corresponding to the 
region enclosed by the $3\sigma_n$ isophote in $K_s$ band. The 
numbers at top right on each row are $3\sigma_n$ isophote axis 
ratios reported in the 2MASS catalogue.   
\label{8gal_ell}}
\end{figure*}

We next examine the ellipticity of galaxy contours at 30 
levels (equally spaced in area in log scale) which covers 
almost the entire region of each galaxy. We compute the 
ellipticity of three auxiliary ellipses ($E_A$, $E_P$, and 
$E_{\chi}$) as well as the ellipticity given by the scalar 
functional for raw images and then apply smoothing procedure. 
Figure \ref{8gal_ell} shows the ellipticity as a function of 
area $A_s$; the top two rows correspond to the unsmoothed 
images, and the bottom two rows to the smoothed versions.
The ellipticity is plotted as a function of area, $A_S$. 
We find that the discrepancy between four estimates of 
ellipticity increases with the growth of the image size 
corresponding to the decrease in the level. The gradient of
the intensity strongly correlate with the level: the lower the 
level the smaller the gradient (i.e. the greater 
$\sigma_n/\Delta f$ as shown in Fig. \ref{inv_g_gal}). 
Therefore, the effects of noise are stronger for lower levels 
corresponding to larger sizes (in $A_S$) which is in qualitative 
agreement with the visual impression (Fig. \ref{8gal_im}). 
The ellipticity of the images after conntour smoothing is 
plotted in two bottom rows in Fig. \ref{8gal_ell}. We see a 
significant reductionof the discrepancy between different 
estimates. The dashed lines correspond to the measurements of 
ellipticity of the $K_s$ images at about $3\sigma_n$ level 
(see e.g. Jarrett et al. (2000) and also the 2MASS Extended 
Source Full-Resolution Image Server at 
http://irsa.ipac.caltech.edu/applications/2MASS/PubGalPS. 
Our measurements are in excellent agreement with the results 
reported in 2MASS catalogue.

All galaxies except E0 and E1 show varying ellipticity that grow 
from lower values in the interior regions to greater values in the 
outer parts. Comparing Figs. \ref{mfs_8_ellipses} and \ref{8gal_ell}
we conclude that this  growth of ellipticity cannot be explained
by seeing and noise effects that become small for $\log_{10} A_S > 1.8$
and thus suggests that it is a real property of NIR galaxies. 
Note that this trend of ellipticity is in agreement with previous 
optical results (Leach 1981, Peletier et al. 1990). All but E0 and 
E1 images show significantly lower ellipticity than in optics.

\begin{figure*}
\epsscale{1.1}
\plotone{\figdir/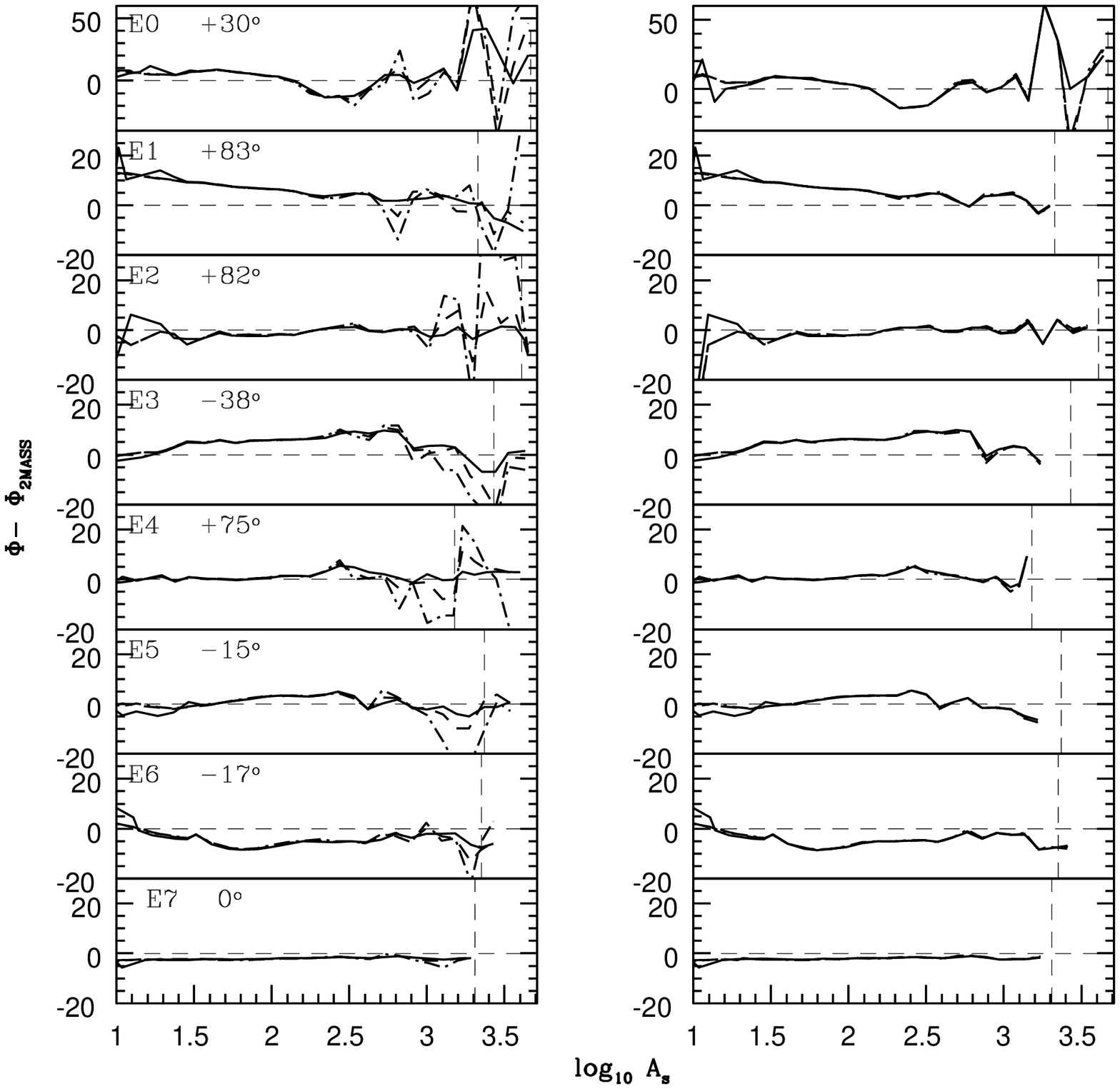}
\caption{The orientations of three auxiliary ellipses as a function
of log$_{10}A_S$. The rows on the left panel show the orientation of 
raw images while those on the right show the orientations after 
smoothing. The horizontal dashed lines show the difference between 
the orientation of each of these ellipses with the 2MASS $3\sigma_n$ 
isophote position angle (shown at the top on the left panel). The 
vertical dashed lines have similar meaning as in  Fig. \ref{8gal_ell}. 
Note that the vertical limit for E0 galaxy is different than all other 
galaxies.\label{8gal_orient}}
\end{figure*}

The orientations of the images are shown in Fig. \ref{8gal_orient} 
as a function of the image size $A_S$. All the rows on the left 
panel show the results before smoothing and the rows 
on the right panel after smoothing. The 2MASS position angles  
measured at $3\sigma_n$ isophote of $K_s$ band images (measured 
east of north) are shown on the left panel. We present the result 
as a difference between our measurement with 2MASS. The horizontal 
dashed lines correspond to no differences. The vertical dashed 
lines show the scalar areas in our measurement corresponding to 
the region enclosed by the $3\sigma_n$ isophote in $K_s$ band. Our 
measurements are in excellent agreement with 2MASS for all galaxies 
at the outer regions except E0. While its orientation is within 
$\sim 10^o$ of 2MASS at the outer part, it varies from $30^o$ 
(where $\log_{10} A_S \sim 2.2-2.4$) to $20^o$ in the central part 
(where $\log_{10} A_S \sim 2.0$). 
It should be kept in mind that a perfectly circular contour has no 
preferred orientation and therefore it is a difficult measure. If 
any circular contour gets perturbed by any kind of disturbance it 
gets deformed from circularity and immediately picks an orientation 
as a consequence of perturbation. 
Since the galaxy contours at different levels are affected 
differently by various spurious effects, we therefore proceed 
with caution regarding the orientation of the E0 galaxy and state 
that the variation in isophote orientation would most probably 
be due to spurious effects and not a true measure. 
While E2, E4, E5, E6, and E7 galaxies show almost stable 
orientations throughout the entire regions (variation is within 
$\pm 4^o$), orientations of E1 and E3 galaxies show interesting 
behavior indicating possible isophotal twists. The E1 orientation 
grows monotonically ($\sim 10^o$) from $\log_{10} \sim 2.7$ towards 
the central part while the E3 changes $\sim 8^o$ around the region 
where $\log_{10} A_S \approx 2.4-2.7$ and gradually decreases 
towards its center.

\subsection{Spiral and Peculiar Galaxies}
\begin{figure*}
\epsscale{2.0}
\plotone{\figdir/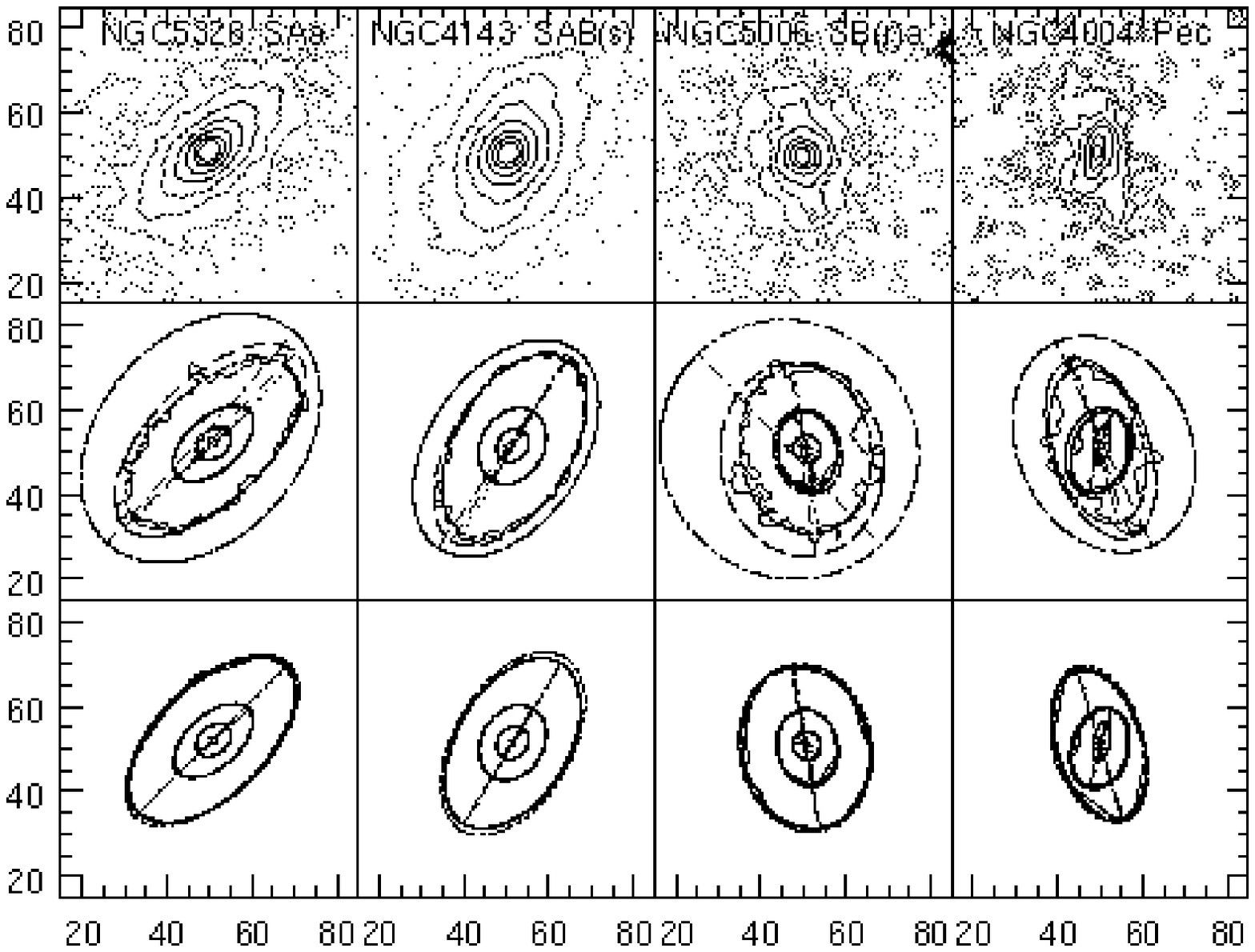}
\caption{The contour plot of four galaxies optically classified as 
spiral and peculiar are shown in the top row. The contours correspond 
to the areas $20\times 2^{n-1}$ in grid units. Three contours with 
areas $A_S \approx$ 50, 200, and 1000 in grid units along with the 
centroids and auxiliary ellipses are shown in the middle row. The 
bottom row shows the the contours after smoothing. The line and 
legend styles are similar as in Fig. \ref{toy_images}. 
\label{4gal_im}}
\end{figure*}

The images of three early-type spiral galaxies NGC5326 (SAa), 
NGC4143 (SAB(s)), NGC5006 (SB(r)a) and one peculiar galaxy 
NGC4004 (Pec) are shown at the top row of Fig. \ref{4gal_im}. 
Three selected contours along with the corresponding auxiliary 
ellipses are shown in the middle (before smoothing) and in the 
bottom row (after smoothing). 
For two of the spirals, NGC5326 and NGC4143, smoothing provides 
both consistent convergence and significant improvement in the 
shape measurements, similar to the spheroidal galaxy cases. The 
other two spirals, N5006 and N4004, however, show remarkable 
inconsistency in the elliptic shape between the inner and outer 
contours (especially for the case of the scalar functional), 
suggesting that their projected shape is not simply an ellipse.

The galaxies NGC5326 and NGC4143 have elliptic contours 
ranging approximately from E2 in the inner parts to E5 in 
the outer parts. The curves are similar to the E5-E6 images 
of the elliptic galaxies as in Fig. \ref{8gal_ell}. 
This actually corresponds well to the visual 
impression: NGC5326 and NGC4143 does look like ellipticals
in $K_s$ band (Fig. \ref{4gal_im} top row). We present the 
orientations of these galaxies at the bottom two rows in this 
figure. The position angle of NGC5326 remains almost stable 
(within $\sim 4^o$) throughout the entire region of the galaxy 
while the orientation of NGC4143 shows some twist ($\sim 10^o$). 
The 2MASS $3\sigma_n$ angles for these two galaxies 
are higher than our measurement: for NGC5326 2MASS  
overestimates by $\sim 4^o$ and for NGC4143 by $\sim 8^o$.

The galaxies NGC5006 and NGC4004 show very distinctive 
behavior. First the ellipticities of these 
galaxies go up and down a couple of times and the estimate 
of the ellipticity based on the scalar functionals is 
significantly higher than that of tensor functionals. 
As we discussed in Sec. 5 the scalar functional 
coincide with the tensorial estimates for ellipses with 
sizes greater than about $\log_{10}A_S \approx 1.8$ and 
ellipticities greater than about E3. The scalar ellipses 
are significantly more elongated than auxiliary ones in 
the ranges $2 < \log_{10}A_S <3$ and E $>$ 3, where they 
must coincide with the other estimates if the images were 
true ellipses (see Fig. \ref{mfs_8_ellipses}). We conclude 
that the discrepancy is real and these parameters can detect 
elliptic  shapes of these images that appear different than 
previous results. 

\begin{figure*}
\epsscale{1.1}
\plotone{\figdir/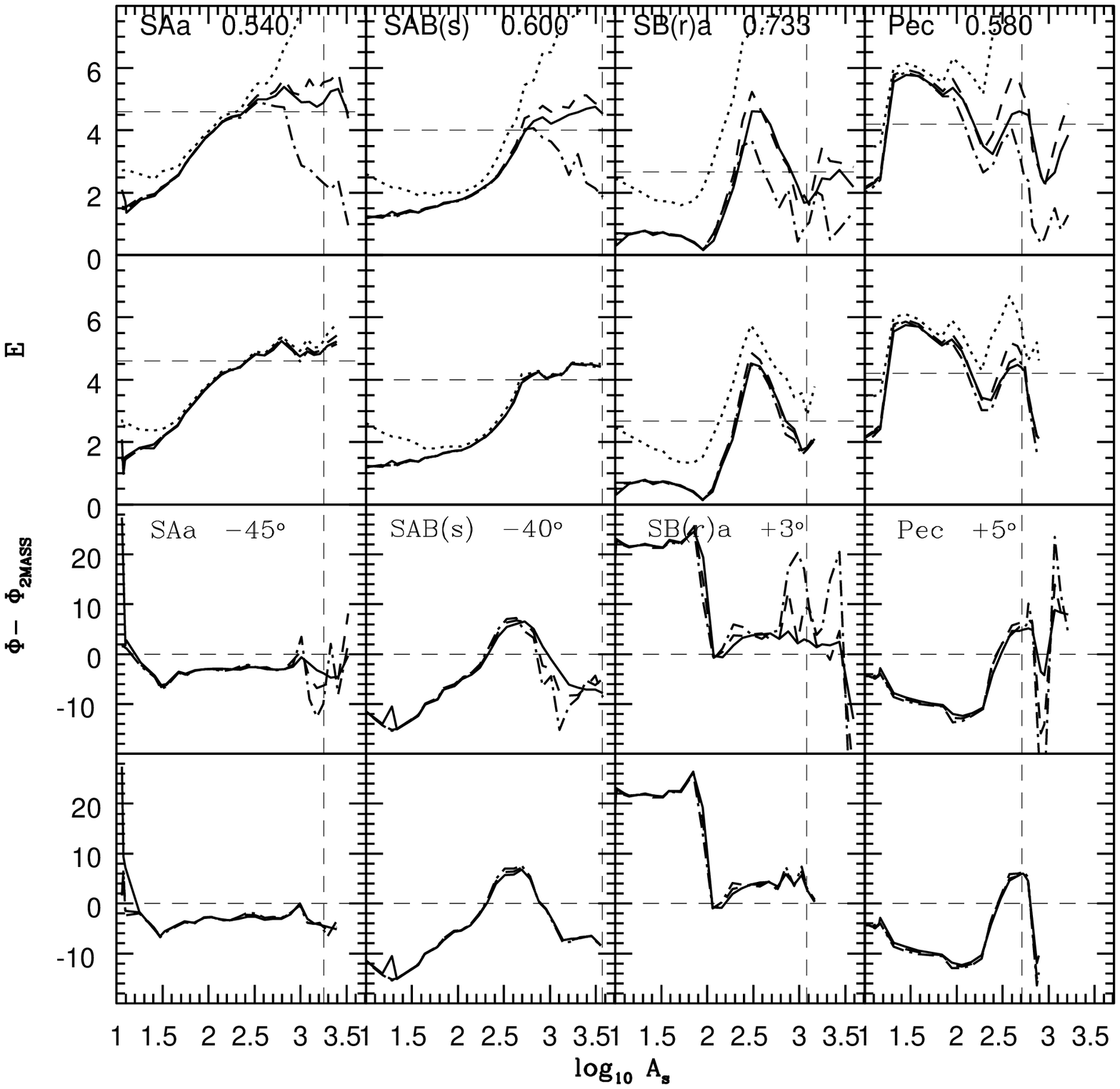}
\caption{The ellipticity of the four galaxies from Fig. 
\ref{4gal_im} are shown as a function of $\log_{10} A_S$ before 
(top row) and after smoothing (second row). The numbers 
at the top row are $3\sigma_n$ isophote axis ratios. The orientation 
of the auxiliary ellipses are shown in the two bottom rows: before 
(second row from the bottom) and after smoothing 
(the bottom row). The $3\sigma_n$ isophote orientations are shown at 
the row second from the bottom. The horizontal dashed lines 
in the top two rows and the vertical dashed lines in all rows have 
similar meaning as in Fig.\ref{8gal_ell}. The horizontal dashed lines 
at the bottom two rows have similar meaning as in Fig. \ref{8gal_orient}.
Note that the differences in orientation for SB(r)a and Pec galaxies 
has been reduced by a factor of 2 for demonstration.
\label{4gal_ell_orient}}
\end{figure*}

The change in orientation of the auxiliary ellipse is also much 
stronger than in the case of elliptic galaxies. Both of 
these galaxies show huge variations in orientation ($\sim 40^o$) 
towards the central region. This change in the direction of 
isophote might be used as a probe to isolate early types galaxies 
from late types. Note that the orientations of these galaxies 
are overestimated by our measurement compare to $3\sigma_n$ angles 
of the 2MASS: $\sim 3^o$ for NGC5006 and $\sim 5^o$ for NGC4004. 
We show the differences in orientations for these galaxies in 
Fig. \ref{4gal_ell_orient} after reducing by a factor of 2. 
It is remarkable that two simple characteristics the ellipticities 
and the orientations of the auxiliary ellipses measured as a 
function of the image size are able to detect varied elliptical 
shapes. We will elaborate this issue in a forthcoming paper.  

\section{Summary}
We have derived and tested a set of morphological parameters 
that can be used for quantification of the geometry and topology 
of two-dimensional images such as galaxies, clusters, and 
superclusters of galaxies. We have demonstrated their efficiency 
by measuring the ellipticity and orientations of hundreds of 
simulated distributions and a dozen of 2MASS images.  

We begin with the construction of a set of contours of constant 
intensity at chosen levels for an image given as a pixelized map. 
To build a contour we use a linear interpolation scheme as 
described in \cite{shandarin-etal02}. Then we compute three scalar, 
three vector, and three tensor (rank 2) \mf.
Three scalar MFs are the area ($A_S$),  perimeter ($P_S$), and 
Euler characteristic ($\chi$) of the region within the contour. 
Two vector MFs $A_i$ and $P_i$ define two centroids which are the 
center of area of the region and the center of mass of the contour 
assuming that the linear density of the boundary are uniform. 
The third vector functional $\chi_i$ defines the center of mass 
of the contour with the linear mass density proportional to the 
curvature of the contour.
In addition, we compute three tensors of the second rank which 
are analogous to the inertia tensors: two assume that the surface 
density of the area and the linear density of the boundary are 
uniform and the third one assumes that the boundary density is 
proportional to thecurvature of the boundary \cite{beisbart00}.

We transform the measured vector and tensor components into 
parameters of three auxiliary ellipses: one having exactly the 
same vector and tensor MFs as the region within the contour 
(eqs. \ref{vA} and \ref{tA}), another as the uniform contour 
(eqs. \ref{vP} and \ref{tP}) and the third one as the contour 
weighted at every point by its curvature 
(eqs. \ref{vchi} and \ref{tchi}). The vector MFs ($A_i, P_i$, 
and $\chi _i$), the corresponding region centroids, also 
represent the corresponding auxiliary ellipse centroids. 
This transformation conserves morphological information and 
provides more homogeneous set of characteristics: four areas 
($A_S, A_A, A_P, A_{\chi}$) and 
four perimeters ($P_S, P_A, P_P, P_{\chi}$), corresponding to 
the contour itself and to the auxiliary ellipses. Finally, the 
MFs provide the position angle directions of the corresponding 
auxiliary ellipses. 
One useful property of these parameterization is that in the 
case of a true elliptic region all areas are the same and 
equals the area of the region itself. Same is true for the 
perimeters and the axes. 

Here we studied in detail the ability of the method to measure
the ellipticities and orientations of contours.
We have tested the effects of grid, atmospheric seeing, and noise
on the described morphological parameters of hundreds of simulated 
elliptic profiles from E0 to E7. We find that for real data, 
represented by $K_s$-band imaging of galaxies from 2MASS, the 
background noise is the dominant factor. In order to reduce it we 
introduced a simple iterative technique for smoothing the perturbed 
contours and tested it on a set of elliptic profiles. We show that 
our method of contour smoothing greatly reduces the effect of image 
noise, allowing accurate and rapid convergence of the shape 
measurements.

As an illustration, we applied the technique to eight images of
elliptic (NGC5044, NGC5328, NGC3608, NGC3091, NGC4742, NGC4008, 
NGC5791, and NGC4550), three spiral (NGC5326 (SAa), NGC4143 (SAB(s)), 
NGC5006 (SB(r)a)) and  one peculiar/irregular (NGC4004(Pec)) galaxy 
taken from the 2MASS catalogue.

We investigated the performance of the method with surface brightness, 
ranging from the inner nuclear regions out to the faintest contours
(Fig. \ref{8gal_ell}, \ref{8gal_orient}, and  \ref{4gal_ell_orient}. 
We show that the ellipticities of all but NGC5044 and NGC5328 
galaxies grow with the size and reach the ellipticity reported in 
Jarrett (2000) and 2MASS catalogue. The measured elliptic shape 
(ellipticity and orientation) is consistent for all four MF 
derivations, unmistakably proving that the projected shape of 
these galaxies is elliptic as seen in the 2-micron window.

Although our sample is small, the results for the asymmetric galaxies 
(NGC5006, NGC4004) in comparison with the spheroidal and normal disk 
spirals, suggest that our method can be used as a powerful discriminant 
of normal, smooth galaxies and those possessing large scale structures 
and asymmetries (e.g., bulges, bars, disk warps, rings, one-armed 
spirals, etc). 
The four ellipticities derived from the MFs, as a function of size, 
show quite different patterns from elliptic galaxies. Combined with 
the position angles they clearly distinguish non-elliptic images. 
We would like to stress that the main goal of this 
study was the measurement of ellipticities and orientations of elliptic 
galaxies. We will report more on the discriminating power of this 
technique in the forthcoming papers. The technique is computationally 
very efficient \cite{shandarin-etal02} and can be applied to the 
analysis of large data sets like SDSS. We shall report the results of a 
larger sample of 2MASS galaxies in the following paper \cite{rah-sh03}.

\noindent{\bf Acknowledgments:}
We are thankful to our referee Tom Jarrett for his valuable comments 
and suggestions. We are also thankful to Robert Cutri and Chris 
Miller for useful communication. NR thanks to Barbara Twarog, Bruce 
Twarog, and Hume Feldman for valuable discussion. SFS acknowledges 
the useful discussion during the workshop at the Aspen Center for 
Physics in June 2002. We acknowledge the use of 2MASS galaxy images 
from 2MASS data archive and also acknowledge the support of the GRF 
2002 grant at the University of Kansas.

\label{lastpage}
\end{document}